\documentclass[usenatbib,useAMS,usegraphicx]{mn2e}

\usepackage{times}

\usepackage{xspace}
\usepackage[fleqn]{amsmath}
\usepackage[rightcaption]{sidecap}
\usepackage{comment}

\providecommand{\eg    }{e.g.\xspace}%
\providecommand{\ie    }{i.e.\xspace}
\providecommand{\xray  }{X-ray\xspace}%
\providecommand{\gray  }{$\gamma$-ray\xspace}%
\providecommand{\grays }{$\gamma$-rays\xspace}%
\providecommand{\fermi }{\textit{Fermi}\xspace}%
\providecommand{\swift}{\textit{Swift}\xspace}%
\providecommand{\planck}{\textit{Planck}\xspace}%
\providecommand{\wise}{\textit{WISE}\xspace}%
\providecommand{\herschel}{\textit{Herschel}\xspace}%

\usepackage{colordvi}
\title[Magnetic Field Amplification and FSRQs]
{Magnetic Field Amplification and Flat Spectrum Radio Quasars}
\author[X. Chen et al.]{%
\parbox{\textwidth}{Xuhui~Chen$^{1,2}$\thanks{chenxuhui.phys@gmail.com},
Ritaban~Chatterjee$^3$,
Haocheng~Zhang$^{4,5}$,
Martin~Pohl$^{1,2}$,
Giovanni~Fossati$^6$,
Markus~B\"ottcher$^{7,4}$,
Charles~D.~Bailyn$^8$,
Erin~W.~Bonning$^9$,
Michelle~Buxton$^8$,
Paolo~Coppi$^8$,
Jedidah~Isler$^8$,
Laura~Maraschi$^{10}$,
Meg~Urry$^{11}$}\vspace{0.1cm}\\
$^1$ Institute of Physics and Astronomy, University of Potsdam, 14476 Potsdam-Golm, Germany\\
$^2$ DESY, Platanenallee 6, 15738 Zeuthen, Germany\\
$^3$ Department of Physics, Presidency University, 86/1, College street, Kolkata-700073, India\\
$^4$ Astrophysical Institute, Department of Physics and Astronomy, Ohio University, Athens, OH 45701, USA\\
$^5$ Theoretical Division, Los Alamos National Laboratory, Los Alamos, NM 87545\\
$^6$ Department of Physics and Astronomy, Rice University, Houston, TX 77005, USA\\
$^7$ Centre for Space Research, North-West University, Potchefstroom 2520, South Africa\\
$^8$ Department of Astronomy, Yale University, PO Box 208101, New Haven, CT 06520-8101, USA\\
$^9$ Department of Physics, Emory University, Atlanta, GA 30322, USA\\
$^{10}$ INAF–Osservatorio Astronomico di Brera, V. Brera 28, I-20100 Milano, Italy\\
$^{11}$ Department of Physics and Yale Center for Astronomy and Astrophysics, Yale University, PO Box 208121, New Haven, CT 06520-8121, USA
      }

\begin{document}
\date{Received 2013; in original form}

%\pagerange{\pageref{firstpage}--\pageref{lastpage}} \pubyear{2013}

\maketitle

\label{firstpage}

\begin{abstract}
We perform time-dependent, spatially-resolved simulations of blazar emission to evaluate several flaring scenarios related to
magnetic-field amplification and enhanced particle acceleration. 
The code explicitly accounts for light-travel-time effects and is applied to flares observed in 
the flat spectrum radio quasar (FSRQ) PKS 0208-512, which show optical/\gray correlation at some times, but
orphan optical flares at other times. Changes in both the magnetic field and 
the particle acceleration efficiency are explored as causes of flares. Generally, external Compton emission appears to describe the available data better than a synchrotron self-Compton scenario, and in particular orphan optical flares are difficult to produce in the SSC framework. 
\xray soft-excesses, \gray spectral hardening, and the detections at very high energies of certain FSRQs during 
flares find natural explanations in the EC scenario with particle acceleration change. Likewise,
optical flares with/without \gray counterparts can be explained by 
different allocations of energy between the magnetization and particle acceleration, 
which may be related to the orientation of the magnetic field relative to the jet flow.
We also calculate the degree of linear polarization and polarization angle as 
a function of time for a jet with helical magnetic field. 
Tightening of the magnetic helix immediately
downstream of the jet perturbations, where flares occur, can be sufficient to explain the increases in
the degree of polarization and a rotation by $\geq 180^\circ$ of the observed polarization angle, if light-travel-time effects are properly considered.
\end{abstract}
\section{Introduction}
\label{intro}
As an extreme class of Active Galactic Nuclei (AGNs), blazars are known to emit electromagnetic waves 
in almost all frequencies that are currently being observed, 
extending from radio to \gray. They are also famous
for being highly variable in an unpredictable manner. An interesting question
is whether blazar variations at different wavelengths are correlated, and if so, how. For example,
whether there are any time lags, what the amplitude relations are. 
The answers to these questions can, e.g., identify the location and mechanism of emission. For one type of blazars, namely,
FSRQs, the correlation between optical emission and GeV \grays is particularly interesting. 
These two energy bands represent the energies either at or beyond the peaks of the two components of 
their spectral energy distributions \citep[SEDs,][]{sambruna_maraschi_urry:1996:seds}. 
The photons in these energies are probably emitted 
by the most energetic particles, and hence exhibit most violent variations. 
Identification of these correlations became possible following the
launch of \fermi and the implementation of its supporting optical monitoring programs, 
such as the Yale/SMARTS program.

\begin{figure*}
\centering
%\vspace*{2cm}
%\floatbox[{\capbeside\thisfloatsetup{capbesideposition={right,top},capbesidewidth=0.33\linewidth}}]{figure}[\linewidth]
%{\caption{A test figure with its caption side by side}\label{fig:obs_lc}}
\includegraphics[width=0.33\linewidth]{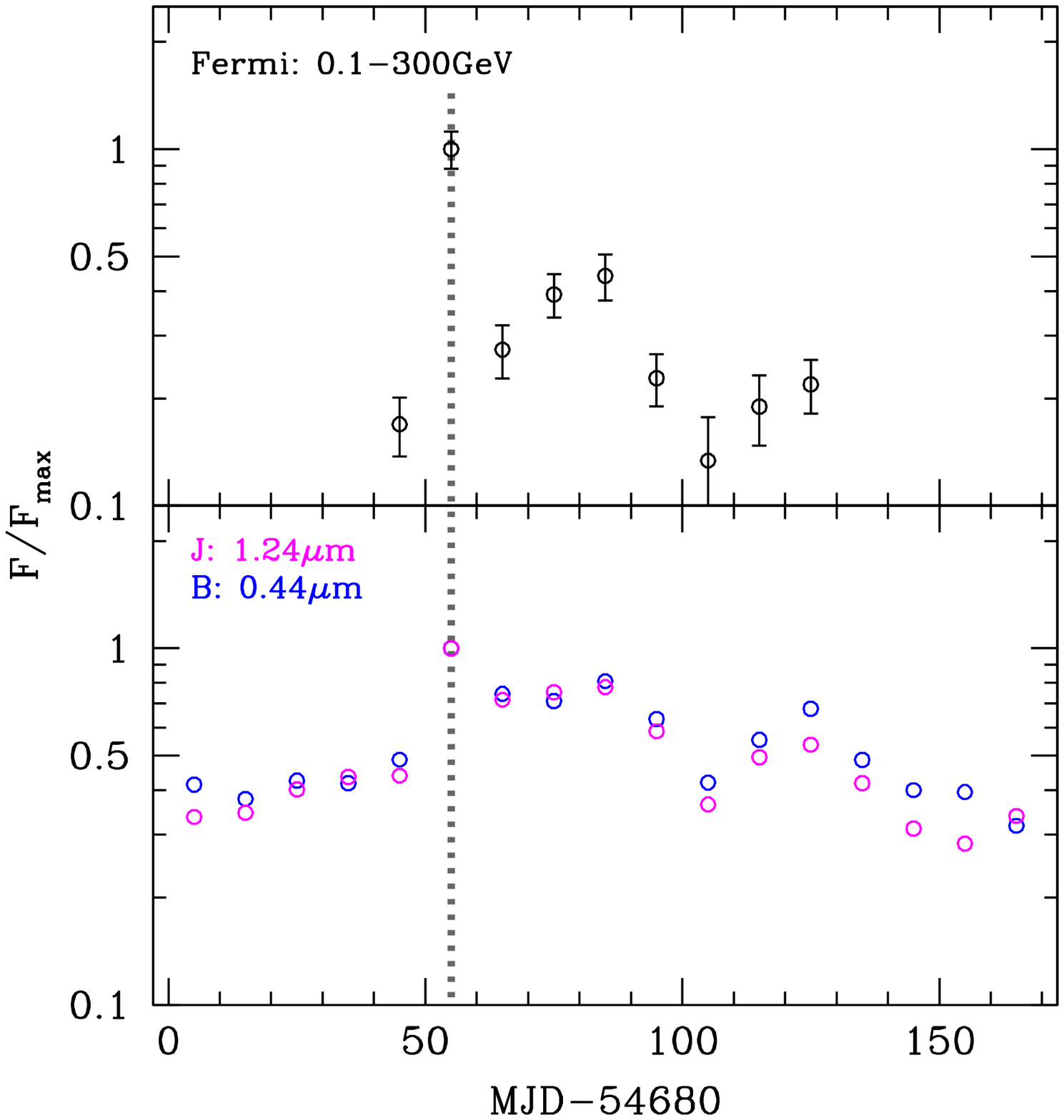}
\hfill
\includegraphics[width=0.33\linewidth]{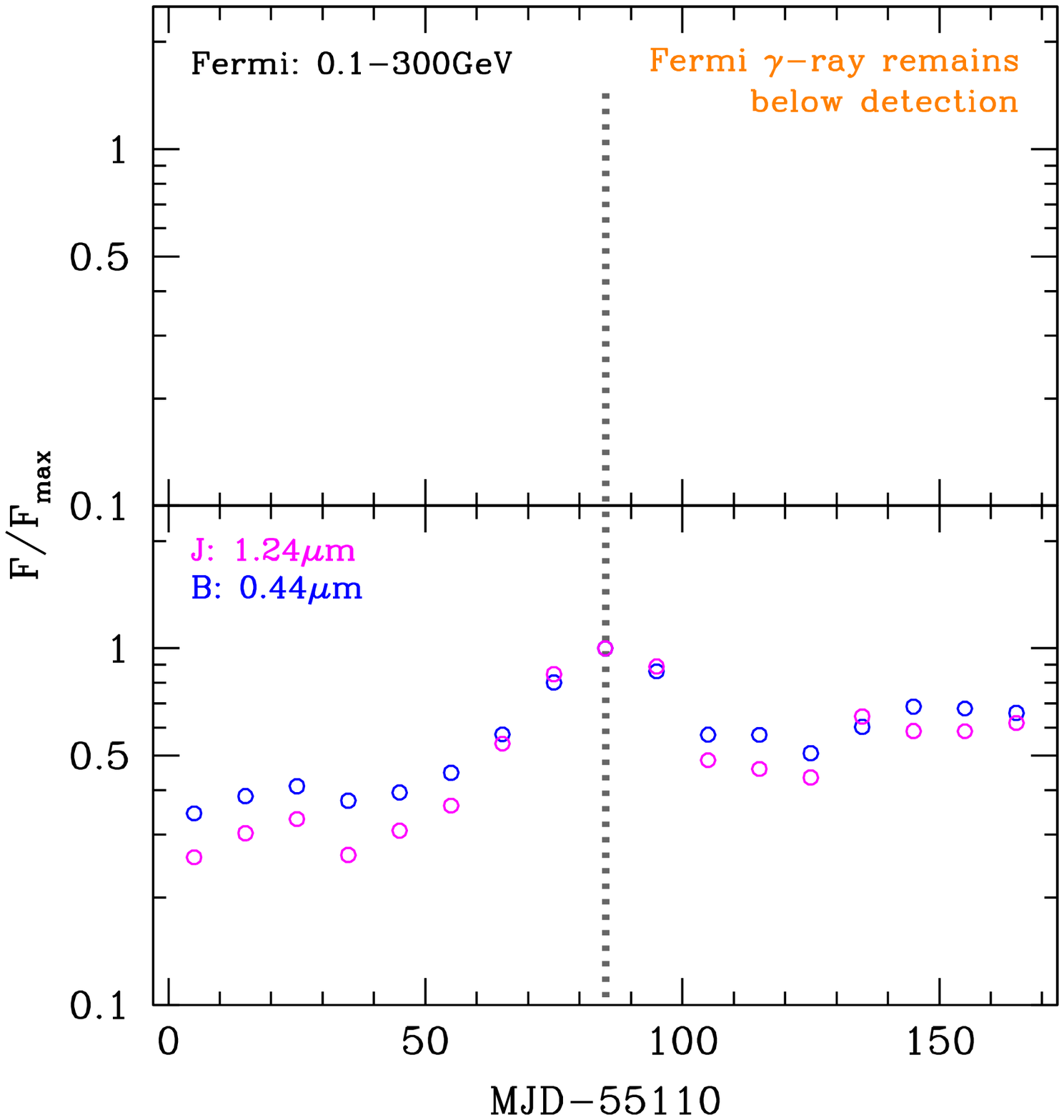}
\hfill
\includegraphics[width=0.33\linewidth]{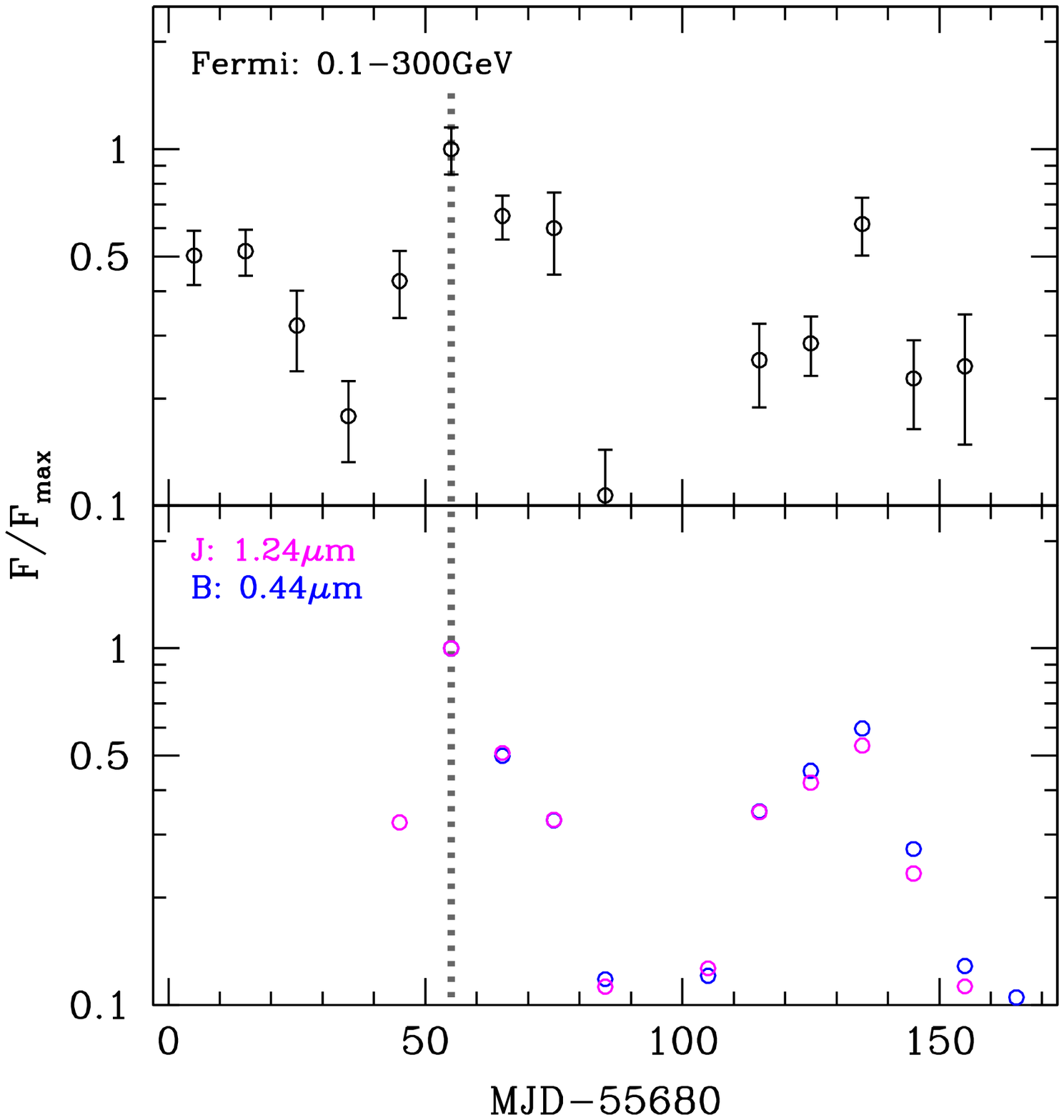}
\caption{\fermi (upper panels) and SMARTS (lower panels) optical/near-infrared light curves in 10-day bins. From left to right, the figures
cover flares 1, 2 and 3 identified by \citet{chatterjee_2013:0208_opticalonly}, starting from
Modified Julian Date (MJD) 54680, 55110, and 55680. The flare peaks are marked with dotted lines.
}\label{fig:obs_lc}

\end{figure*}

In most cases the correlation between these two bands are established 
\citep{bonning_2012:smarts_blazar,chatterjee_2012:smarts_similarity}.
However, \citet{chatterjee_2013:0208_opticalonly} identified at least one case of such correlation breaking
down \citep[see also Fig.\,2 of][]{hess_2013:1510}. 
In contrast to the "orphan" \gray flares occasionally found in BL Lac objects 
\citep{krawczynski_etal:2004:1es1959},
these FSRQs show strong optical flares without \gray counterparts. 
The authors of \citet{chatterjee_2013:0208_opticalonly} identified
three major optical flares from PKS 0208-512 over a period of 3 years.
\gray activity in flares 1 and 3 is highly correlated with that at the optical band.
But in flare 2, the \gray flux remains at a low level. 
The question arises why the same source exhibits correlated optical/\gray flares sometimes, but
orphan optical flares at other times.

Since the optical emission is generally accepted to be
produced by synchrotron emission \citep{bregman_1988:synchro:293.714,urry_mushotzky:1982,
impey_neugebauer88,marscher_1998:144.25},
it is sensible to postulate that if the flare is caused by a change of
the magnetic field, it may not have a direct effect on the \gray emission,
except the indirect effect through particle cooling and acceleration.
It is known that magnetic field can be amplified by astrophysical shocks beyond simple
shock compression through a turbulent dynamo effect. This has been proved both numerically \citep{guo_2012:mhd_bampli}
and analytically \citep{fraschetti_2013:analytical_bampli}, 
and has been applied to explain
observations of supernova remnants \citep[SNRs,][]{parizot_2006:snr_bampli} 
and \gray bursts \citep[GRBs,][]{mizuno_2011:mhd_bampli}.
If this kind of amplification is also at work in relativistic jets 
which presumably harbor blazar flares, it can be expected to explain the orphan optical flares.

\citet{chatterjee_2013:0208_implication} studied these anomalous flares of PKS 0208-512 
with a time-independent model. However, since blazar flares are naturally time-dependent
phenomena, it is important to account for the timing information with time-dependent modeling.
We will briefly describe the data reduction and analysis in $\S$ \ref{observe}.
The comparison between time-dependent modeling and the observation 
are presented in $\S$ \ref{results}, followed by discussion in $\S$ \ref{discussion}.

\section{Multiwavelength data of PKS 0208-512}
\label{observe}

PKS 0208-512 is a high-power blazar with redshift of 1.003. 
It has been classified as a BL Lac object based on its line properties \citep{healey_etal:2008:cgrabs}. 
However, \citet{gg_2011:bllac_fsrq} noted that the broad-line emission from PKS 0208-512 is in fact strong ($L_\mathrm{MgII}\sim10^{44}\textrm{erg s}^{-1}$), 
on top of the fact that the SED of this source resembles those of FSRQs. 
So it is suggested that this source should be re-classified as an FSRQ.

We obtained spectra of PKS 0208-512 at the MeV-GeV energy range by analyzing data from \textit{Fermi}/LAT. 
For this purpose, we used the standard \textit{Fermi}/LAT Science Tools software package (version v9r27p1). 
We analyzed a Region of Interest of 15$^\circ$ in radius, centered at the position of PKS 0208-512, 
using the maximum likelihood algorithm implemented in \texttt{gtlike}. 
We included all sources within 15$^\circ$ of PKS 0208-512, extracted from the \textit{Fermi} 2-yr catalog (2FGL), 
with their normalizations kept free and spectral indices fixed to their catalog values. 
We use the data selection P7SOURCE\_V6 and its response functions, the diffuse background model \texttt{gal\_2yearp7v6\_v0.fits}, 
and the isotropic background model \texttt{iso\_p7v6source.txt}. 
To calculate the spectra, we divide the spectral range 0.1-24.3 GeV into two spectral bins and model PKS 0208-512 
with a simple power law at each spectral bin with the spectral index and normalization kept free. 
We obtained the \gray spectra at various 10-day intervals during flares 1 and 3 
\citep[as defined in][]{chatterjee_2013:0208_opticalonly} in order to probe the 
spectral changes during the rise and decay of those outbursts.
Smaller time intervals can not always give two spectral points with enough statistical significance, 
while larger time intervals can not resolve the profile of the flares.

\begin{figure*}
\centering
\includegraphics[width=0.45\linewidth]{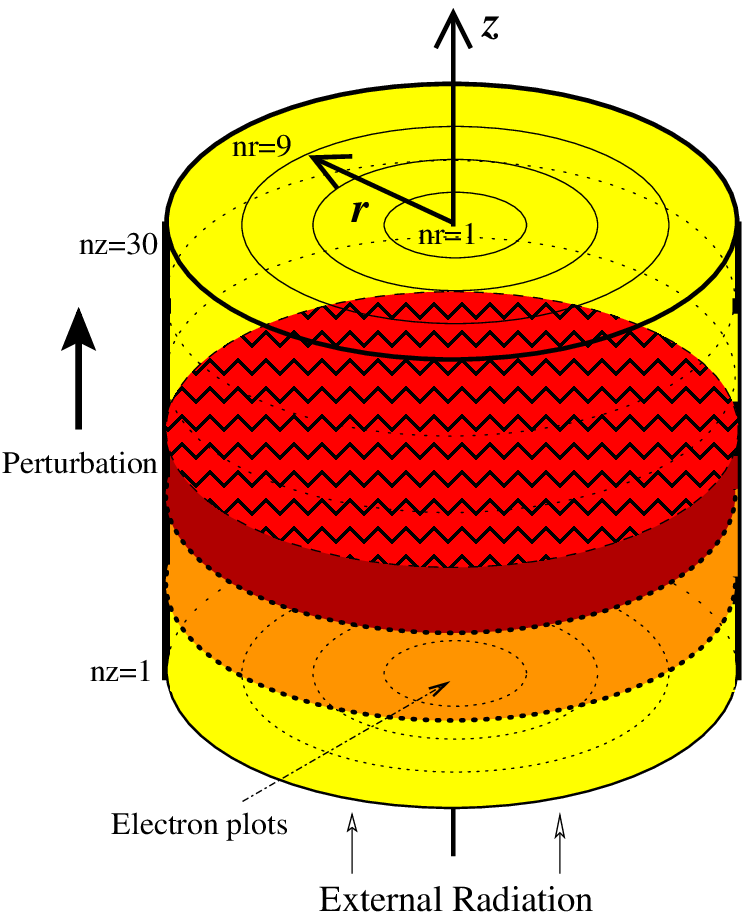}
\hfill
\includegraphics[width=0.45\linewidth]{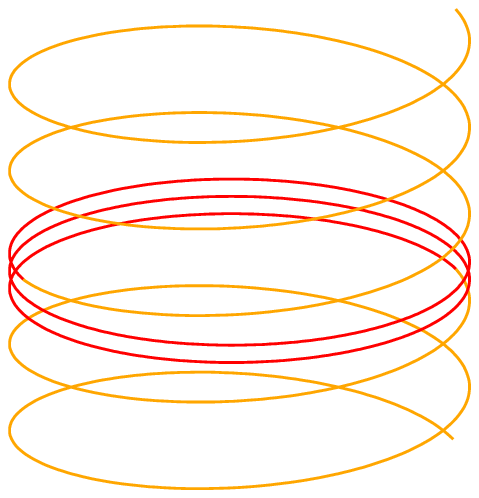}
\caption{{\bf Left}: Sketch of the model geometry and its
co-moving 2-D grid with azimuthal symmetry.  
The hatched layer represents a perturbation that is stationary in the black-hole frame and
relativistically moving upward in the blob rest frame. The yellow zones represent quiescent regions in which the particles have reached equilibrium between acceleration and losses.
We plot in red the zones in which the magnetic field or the acceleration efficiency have changed to induce a flare. The orange zones represent the regions in which the gain and loss rates  are back to their baseline levels, but particle spectra are still elevated.
{\bf Right}: A sketch of the helical magnetic field used in \S \ref{ec:theta}. The red portion of the field line
stands for the perturbed region, where an enhancement of the azimuthal component of the field leads
to the tightening of the helix, and hence a change in the field direction.
}
\label{fig:geo}
\end{figure*}

Light curves in the $B$, $V$, $R$, $J$, and $K$ band are from the Yale/SMARTS optical--near IR blazar monitoring program\footnote{http://www.astro.yale.edu/smarts/fermi/} 
\citep{ bonning_2012:smarts_blazar,chatterjee_2012:smarts_similarity} 
which uses the ANDICAM instrument on the SMARTS 1.3m telescope located at CTIO, Chile. 
Details of observations and data reduction procedures are described in \citet{bonning_2012:smarts_blazar}. 
We calculate the OIR SEDs and lightcurves by averaging the flux over 10-day intervals.

We obtained the X-ray data from the \textit{Swift}/XRT monitoring program of \textit{Fermi}/LAT sources of interest\footnote{http://www.swift.psu.edu/monitoring/}. 
We use the \emph{Swift}/XRT data products generator\footnote{http://www.swift.ac.uk/user\_objects/docs.php}
\citep{evans_2009:swiftxrt_web}, 
and fitted the data with the \xray spectral fitting package XSPEC.
During the time interval of interest (MJD 55190-55200, high state of flare 2), 
\swift-XRT only observed the source for one 1351-s period on MJD 55195.

The high- and low-state SEDs from flares 1 and 2 are shown with the simulation results in 
Fig.\,\ref{fig:sscb}$-$\ref{fig:ecacc}.
In Fig.\,\ref{fig:obs_lc} we show the light curves of the three flares as identified by
\citet{chatterjee_2013:0208_opticalonly}. 
Note that flare 3 can also be viewed as an ensemble of two flares peaking at MJD 55735 and 55815.
The light curves of flare 1 and 3 in optical and \grays show striking similarity
with no apparent delay between the peaks at different energies.
Flare 2 is not significantly detected in \grays in 10-day bins. We show its optical
light curves together with the simulation results.

\section{The model}
\label{model}

We use the time-dependent multizone blazar model 
built by \citet{chen_etal:2011:multizone_code_mrk421,
chen_etal:2012:424.789} to study the multiwavelength data set of PKS 0208-512. 
This model employs a axisymmetric cylindrical geometry (see Fig.\,\ref{fig:geo}), divided
into many zones in radial and longitudinal directions to account for inhomogeneity.
This inhomogeneous blob, whose emission is assumed to dominate over the continous jet stream 
because of higher density and/or stronger magnetic field,
travels relativistically in the AGN frame and encounters
a stationary perturbation that, for simplicity, is treated as a flat feature. In the blob frame, it is the perturbation that travels through
the blob and causes a change in the plasma condition, hence initiating the flare.
The model is based on our 2-D Monte-Carlo/Fokker-Planck (MCFP) code. The
Monte-Carlo technique is used for
the radiative transfer, so that all light-travel-time effects (LTTEs) are taken into account. 

The temporal evolution of the electron (and positron) population in each simulation zone
is obtained by solving the Fokker-Planck equation for the differential number density of electrons, $N(\gamma,t)$:
\begin{equation}
\label{eq:FPeq}
\begin{split}
 \frac{\partial N (\gamma ,t)}{\partial t} & =
-\frac{\partial}{\partial \gamma}\bigg[N(\gamma,t)\dot{\gamma}(\gamma,t)\bigg] \\
 & +\frac{\partial}{\partial \gamma}\bigg[D(\gamma,t)
 \frac{\partial N(\gamma,t)}{\partial \gamma} \bigg]
 + Q(\gamma,t) - \frac{N(\gamma, t)}{t_\mathrm{esc}}\ .
\end{split}
\end{equation}
On the right hand side of eq.(\ref{eq:FPeq}), the first term describes continuous cooling
and acceleration of the particles with rate $\dot{\gamma}$, 
\begin{equation}
\label{eq:acc}
\dot{\gamma}(\gamma,t)=\dot{\gamma}_\mathrm{cool}(\gamma,t)+\dot{\gamma}_{D}(\gamma,t) \ ,
\end{equation}
which includes stochastic particle acceleration with rate
\begin{equation}
\dot{\gamma}_{D}(\gamma,t)=\frac{2}{\gamma}D(\gamma,t)=\frac{\gamma}{t_\mathrm{acc}}\ ,
\end{equation}
for which 
the acceleration time scale is defined by $t_\mathrm{acc}$.

The second term describes the diffusion of the particle
in momentum space with diffusion coefficient
\begin{equation}
D(\gamma,t)=\frac{\gamma^2}{2 t_\mathrm{acc}}\ .
\end{equation}
The acceleration process is similar to that described by \citet{katarzynski_etal:2006:stochastic}.
It is a result of particle diffusion in momentum space, which mainly represents
the second-order Fermi process. The acceleration time scale is assumed to be
independent of particle energy here. One can also reasonably assume that
the acceleration time scale changes with particle energy, \eg it increases linearly with energy.
The resulted particle spectrum will deviate from a simple power-law. A detailed study
of an energy dependent acceleration will be the subject of a separate publication.

The third term of eq.(\ref{eq:FPeq}), $Q(\gamma,t)$, describes particle injection;
the fourth term accounts for particle escape through an adjustable parameter $t_\mathrm{esc}$ that is not changed throughout the simulation.

The quiescent state of the jet is established through quasi-monoenergetic injection at 
Lorentz factors $\gamma_\mathrm{inj}$ which is technically realized with a narrow Gaussian. Such injection spectra may arise from sweep-up of ambient material by a fast plasma cloud \citep{pohl_2000:blastwave:354.395}, 
in quasi-perpendicular relativistic shocks, 
or through the pile-up effect caused by the balance between Fermi acceleration and radiative cooling
\citep{schlickeiser_1984:pileup:136.227}. 
But here it is merely a parametrization of pre-acceleration throughout the jet with steady-state rate
\begin{equation}
Q(\gamma) = Q_0\,\exp\left(-\frac {(\gamma -\gamma_\mathrm{inj})^2}{2\,\sigma_\gamma^2}\right)\ .
\end{equation}
The pre-accelerated particles are then subjected to stochastic acceleration in the entire simulation box. The comparison between $t_\mathrm{acc}$ and the cooling time scale, which is energy dependent, gives the 
maximum energy of particles; the ratio between $t_\mathrm{acc}$ and $t_\mathrm{esc}$ decides the spectral index
of the particle power-law distribution. Particle acceleration, cooling, injection and escape reach an equilibrium,
which forms the quiescent state of the system, with parameter values as listed in Table~\ref{tab-1}. The simulations need some time to reach equilibrium, and so flares are launched only when equilibrium condition have been reached.

Flares are induced by changing one or more parameters of the system in a thin zone plotted red in Figure~\ref{fig:geo}, that is assumed stationary in the host-galaxy frame and therefore propagates relativistically in the simulation frame. Such standing features can arise from external perturbations or through recollimation. We do not intend to specify the origin of the perturbation, but focus on its effect. In this paper we specifically discuss changes in the magnetic-field and in the acceleration time, $t_\mathrm{acc}$. 
More details about the particle processes in our model can be found in \citet{chen:2012:thesis}.
Example electron distributions in individual zones are shown in Fig.\,\ref{fig:sscb-ele} and the bottom panels of 
Fig.\,\ref{fig:sscacc}-\ref{fig:ect}. 

We set the angle between the line of sight and the jet axis as $\theta=1/\Gamma$, and so the Doppler factor $\delta$
is always equal to the bulk Lorentz factor $\Gamma$.
In the comoving frame of the jet, this is equivalent to observing at 90$^\circ$ to the jet axis.
Except in \S\,\ref{ec:theta}, the magnetic field is assumed to be fully disordered.
The conversion between flux and apparent luminosity is done with the cosmological parameters 
$\Omega_m=0.28$, $\Omega_{\Lambda}=0.72$, $H_0=70$ km/s/Mpc.
A steady thermal component is added to the simulation during post-processing, using
an average thermal-emission component from radio-loud quasars \citep{elvis_etal94_qsr_sed} 
scaled according to the observed \swift/UVOT flux of PKS 0208-512.
Note that this average thermal component is directly deduced from observation, without needing to specify
whether it comes from the accretion disk, \xray corona, or anything else.
All the simulated SEDs aim to match their observational counterparts from flare 2, \ie the filled triangles in the plotted SEDs.
Matching between the observational and simulation light curves is achieved through variation of the starting points,
\ie they can be shifted horizontally to match each other. In Fig.\,\ref{fig:sscb}-Fig.\,\ref{fig:ect},
the SED and light curve plots have the same observational data. 
The pre-flare/flaring SED data are 10-day averages centered on MJD 54725/54735 for flare 1, 
and MJD 55165/55195 for flare 2. The historical data include (grey squares) \swift, \planck and ground based radio data in November of 2009 reported by \citet{giommi_2012:planck}, 
and (grey circles) \wise data taken in MJD 55367-55369 (around June 23, 2010).

\section{Results}
\label{results}
\subsection{Pure SSC scenario}
\label{ssc}
\subsubsection{Brief change of magnetic-field strength}
\label{ssc:B}

\begin{figure}
\centering
\includegraphics[width=0.97\linewidth]{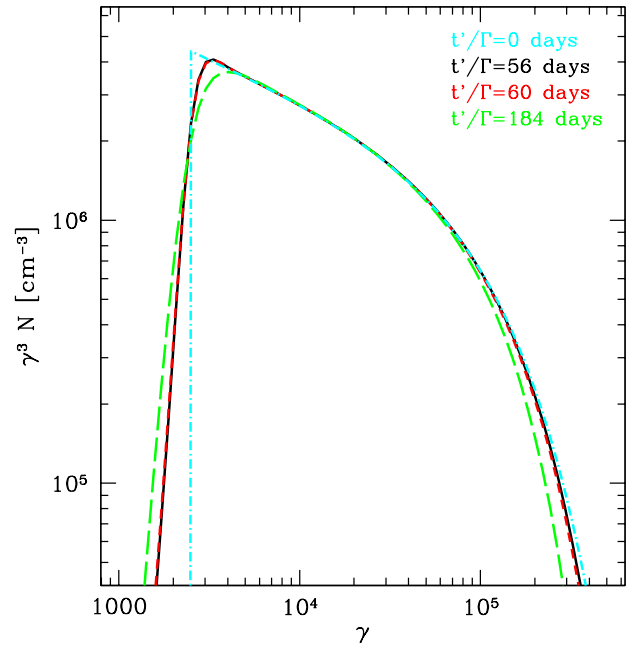}
\caption{The electron spectrum in the bottom-center cell
(nr=1,nz=1 in Fig.\,\ref{fig:geo})
in the pure Synchrotron Self-Compton (SSC) case with
change of the magnetic field as the cause of the flare. 
The red line corresponds to the time when the perturbation has reached the bottom zone, i.e. displays flare spectra.
$t'/\Gamma$ are shown for reference only.
They do not correspond to any specific observer's time $t$ because of the LTTEs.
}
\label{fig:sscb-ele}       % Give a unique label
\end{figure}

We begin the investigation with a pure SSC scenario. In this case, five key parameters (magnetic field $B$, 
electron density $n_{e}$, volume length $Z$ (or radius $R$), beaming Doppler factor $\delta$,
Lorentz factor of the injected low energy particle $\gamma_{inj}$) are constrained
by 5 observables (synchrotron and IC peak frequencies $\nu_{sy}, \nu_{ic}$, synchrotron and IC 
apparent luminosity $L_{sy}, L_{ic}$, and variability time scale $t_{var}$). We note that the symmetry of
most light curves in blazars \citep[similar rise and decay time,][]{chatterjee_2012:smarts_similarity} 
indicates $t_{var}$ is directly linked to LTTEs.
The parameters used are summarized in Table \ref{tab-1}.

\begin{figure}
\centering
\includegraphics[width=0.97\linewidth,clip=]{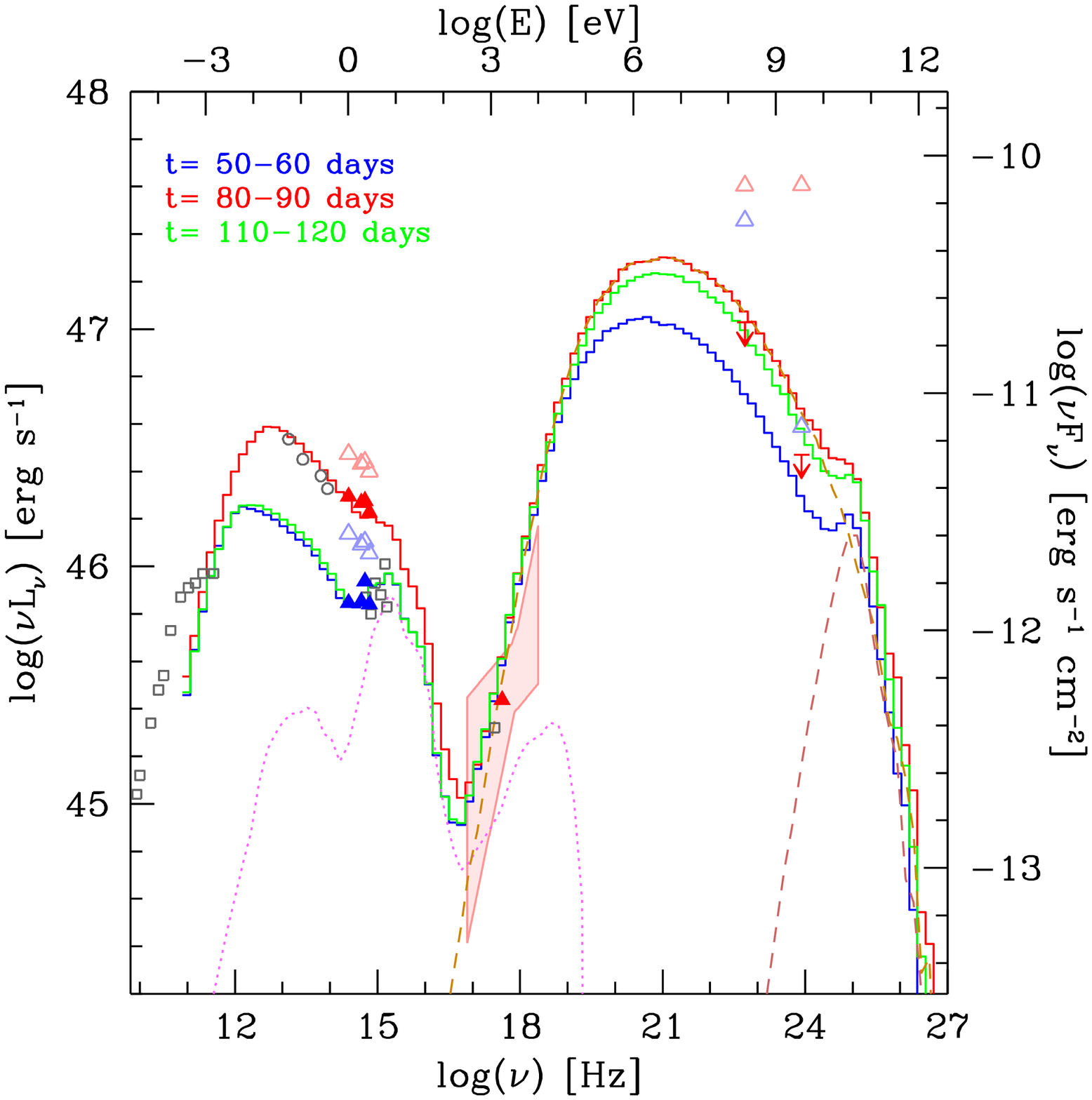}
\includegraphics[width=0.97\linewidth]{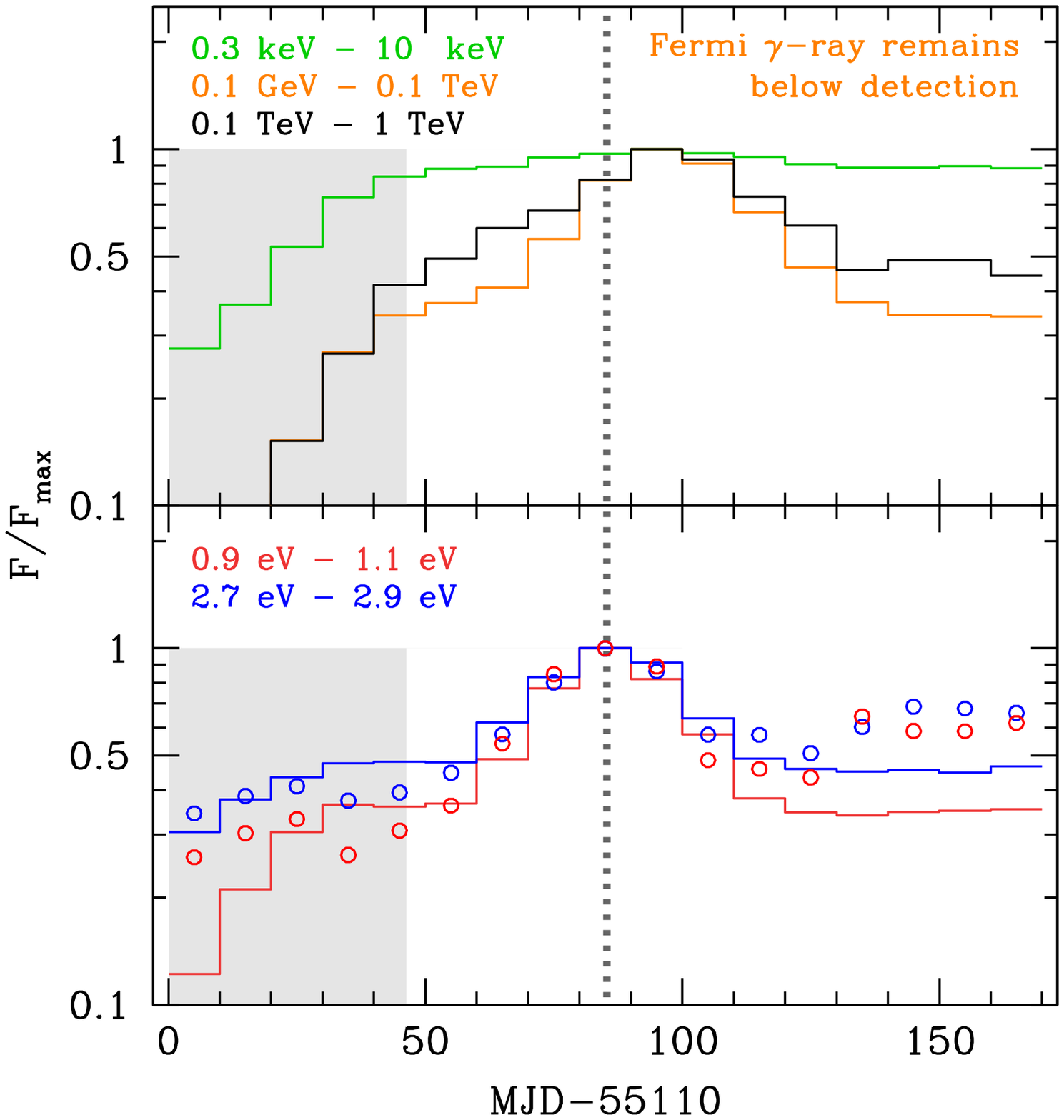}
\caption{The SEDs (top) and light curves (bottom) 
in the pure Synchrotron Self-Compton (SSC) case with
change of the magnetic field as the cause of the flare. 
{\bf Top:} The colored triangles are simultaneous data 
points from SMARTS, \swift and \fermi, 
with blue/red being the pre-flare/flaring states. The grey squares and circles are historical data.
The open blue/red triangles are part of flare 1 identified by \citet{chatterjee_2013:0208_opticalonly},
while the filled blue/red triangles are for flare 2.
The \swift bow-tie and the \fermi upper limits are simultaneous with the filled red triangles.
% The grey data are not strictly simultaneous with the colored data. 
The three histograms show the simulated SEDs before, during, and after the peak of flare 2, and so they should match the filled triangles, 
while other data points are plotted there for reference.
The dotted magenta line is the steady thermal emission component. 
The dashed orange red lines refer to the first second order SSC emission 
during the peak of the flare.
{\bf Bottom:} In the lower panel the open circles show the 10-day-averaged optical light curves 
in B (blue) and J (red) bands during flare 2. 
The histograms show two simulated synchrotron light curves
at similar frequencies. In the upper panel there are no observational data points.
Fermi \gray flux was always below detection in 10-day bins during this time.
Three simulated IC light curves are shown, with green, orange,
and black solid lines representing the energy bands in \xray, \fermi \gray, and very high energy (VHE) \gray.
The shaded grey areas mark the phase when the simulation is still in setup phase. 
The vertical dotted line marks the peak of
the synchrotron flare.}
\label{fig:sscb}       % Give a unique label
\end{figure}

The flare is assumed to be caused by an increase of magnetic-field energy density
(by a factor of 20) immediately downstream of
the stationary perturbation (Fig.\,\ref{fig:geo} left, red cylinder). The thickness of the region with increased magnetic field is $1/10$
of the length of the emitting blob. The same thickness is used in subsequent cases.

The resulting electron distributions are shown in Fig.\,\ref{fig:sscb-ele}, whereas the SEDs and the light curves are displayed in 
Fig.\,\ref{fig:sscb}. Perturbations only in the magnetic field leave 
the electron distribution virtually unchanged during the entire simulation.
Variability in the radiation flux is therefore mostly caused by 
the modification of the emissivities and geometric effects. The blazar flares both in the optical and in $\gamma$ rays, and the \gray
flares seem to be smoother, longer lasting, and, most prominently, appear to be delayed compared to the optical flare. The delay is roughly 10 days, 
which is about 1/5 of the entire flare duration.
Both the delay and smoothing of the \gray light curve are caused by the internal LTTE of the SSC emission. 
The inconsistency between this delay and the lack of time delays in observations such as those shown in 
Fig.\,\ref{fig:obs_lc} indicates that the SSC scenario with magnetic-flux change does not explain
the correlated flares in FSRQs, unless the emission region is much smaller than assumed here, 
and the light curves are symmetric for non-geometric reasons.
Neither does it explain the optical only flares, because the model predicts that the SSC flux
will also increase and violate the \fermi upper limits in the SED.

Another interesting feature in Fig.\,\ref{fig:sscb} is the presence of second-order SSC emission above
tens of GeV. Despite being largely suppressed by Klein-Nishina (KN) effect, the 
second-order SSC component is visible
as an additional bump in the SED.
Since the SEDs of some high redshift blazars such as 
3C 279 \citep{albert_etal:2008:3C279_magic_detection} and PKS 1424+240 \citep{furniss_2013:1424_redshift} 
appear to have an up-curving shape at VHE after
the correction of extragalactic background light (EBL) absorption, 
it is tempting to explain those SEDs with the second order SSC emission.
However, PKS 1424+240 is classified as an intermediate-frequency peaked BL Lac (IBL) or 
high-frequency peaked BL Lac (HBL). The higher peak energy of the synchrotron component
suggests that a significant second-order SSC emission is very unlikely on account of KN suppression.
Since these different types of blazars show similar up-curving SED only after EBL de-absorption,
it may be more natural to explain these curves as a result of the uncertainty in the EBL models used.
\begin{figure}
\centering
%\vspace*{2cm}
\includegraphics[width=1.0\linewidth,height=0.87\linewidth,clip=]{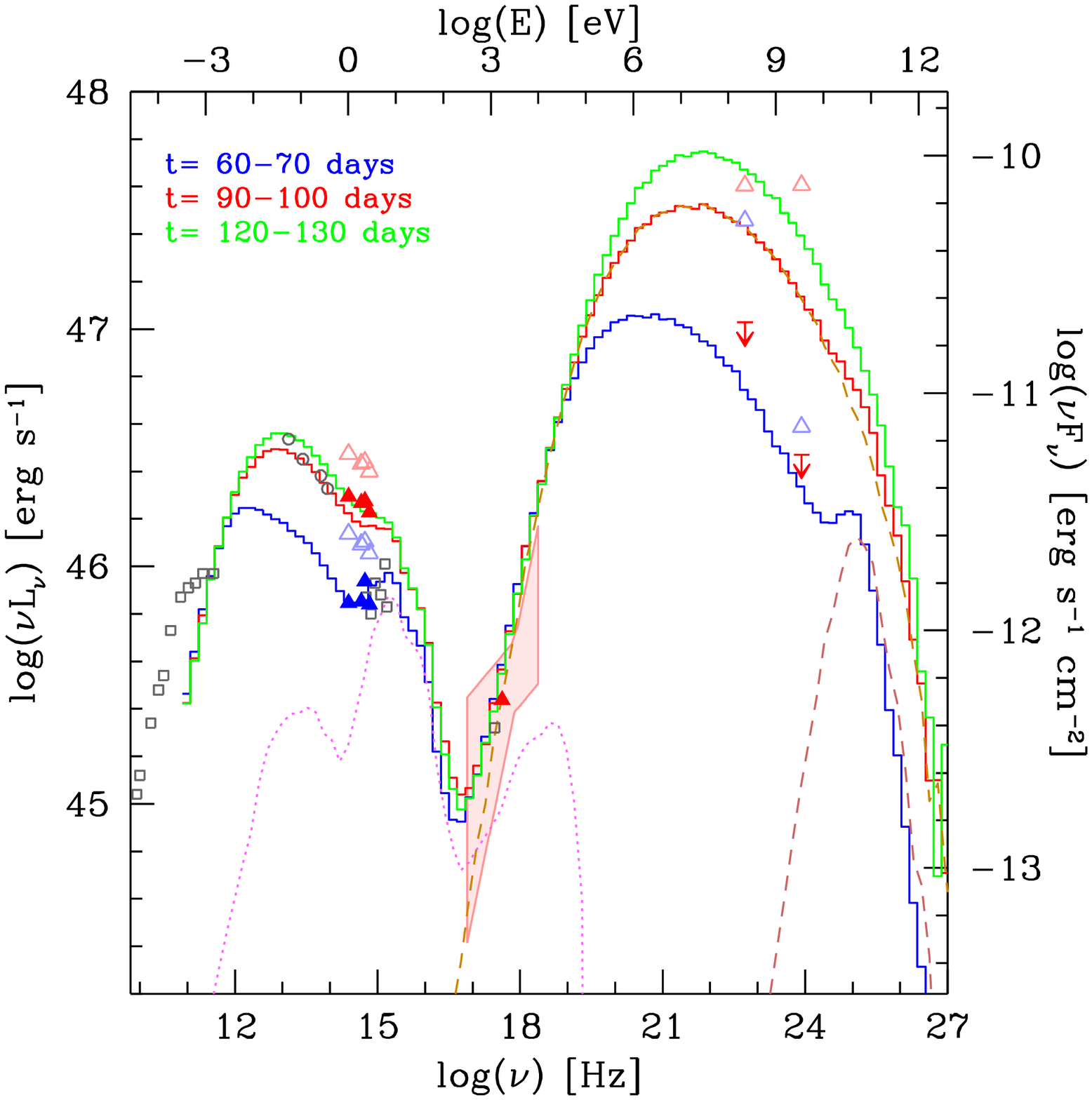}
\hfill
\includegraphics[width=1.0\linewidth,height=0.84\linewidth,clip=]{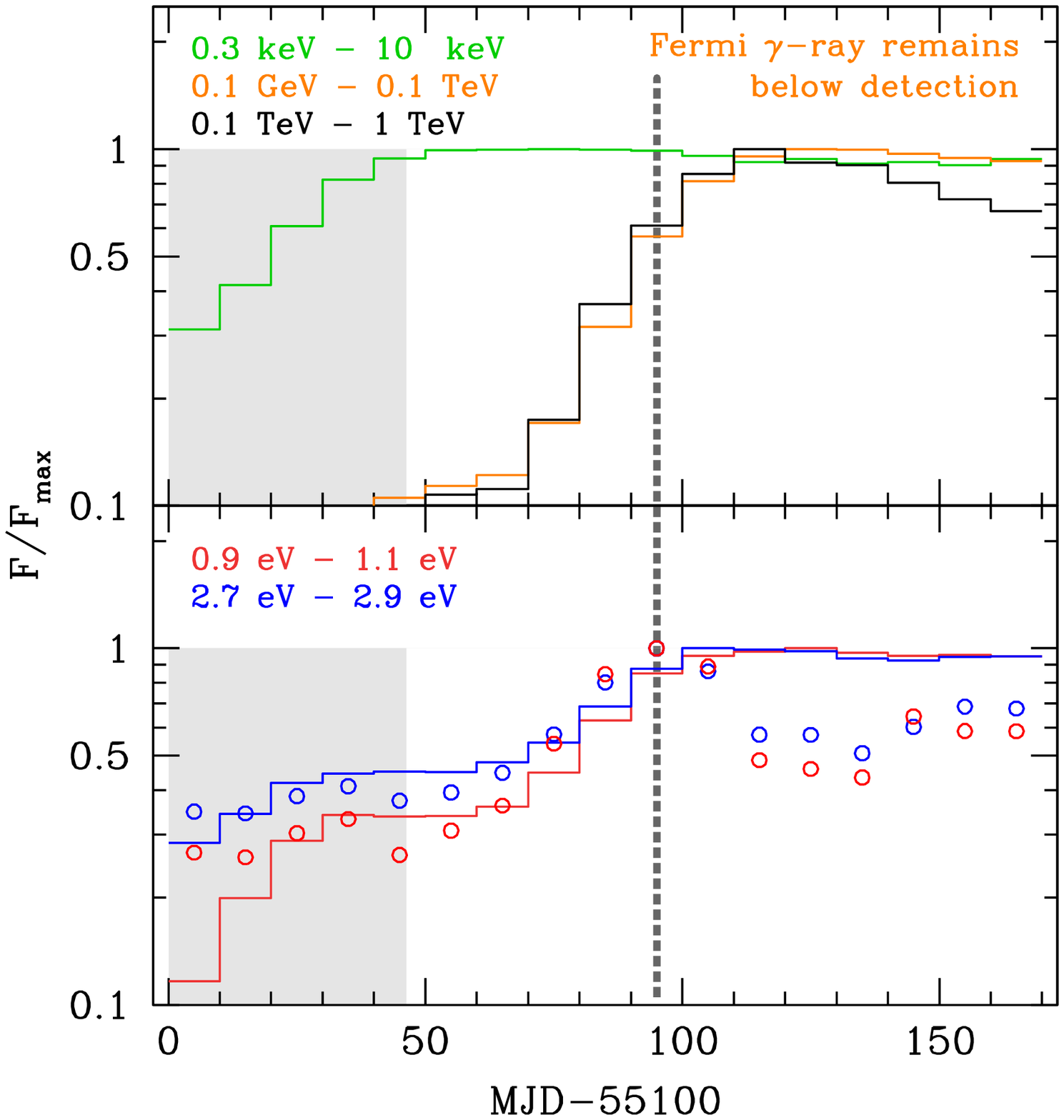}
\hfill
\includegraphics[width=1.0\linewidth,height=0.85\linewidth,clip=]{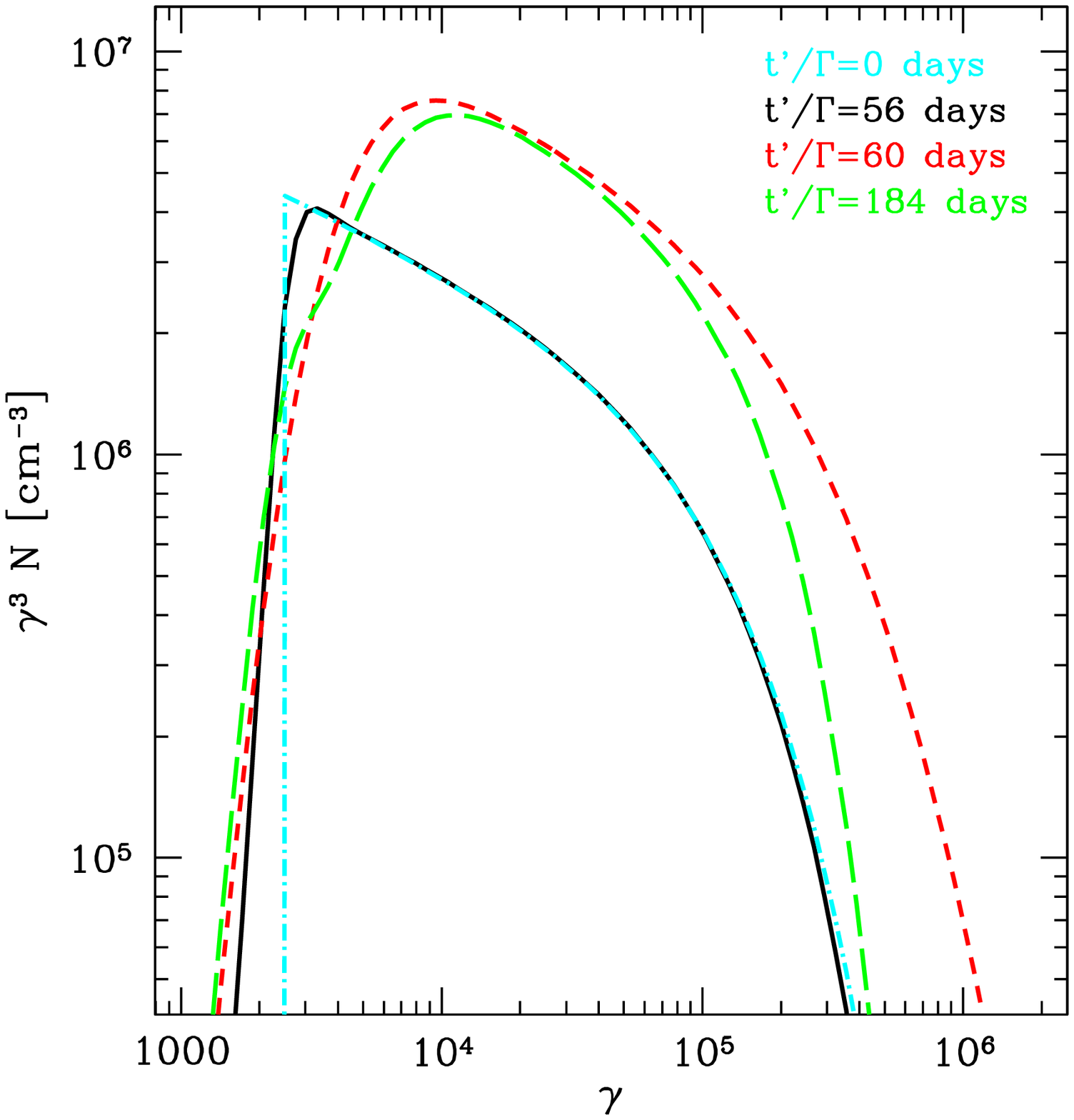}
\hfill
\caption{The SEDs (top), light curves (middle) and electron distributions (bottom) of the SSC 
case with change of particle acceleration efficiency as the cause of the flare. 
The color schemes are similar to those in Fig.\,\ref{fig:sscb}, 
except there is a 10-day shift of the observation 
with respect to the simulation data in order to make them more comparable.}
\label{fig:sscacc}       % Give a unique label
\end{figure}

\subsubsection{Brief change of acceleration efficiency}

Current theories of magnetic-field amplification typically involve the generation of strong turbulence
\citep{bell_2004:bamplification:353.550, mizuno_2011:mhd_bampli}, which
can be expected to cause strong stochastic particle acceleration.
If a flare is caused by a change in the particle distribution, e.g. originating from a change in the
efficiency of particle acceleration, the SSC emission would show an immediate response, long before the effect of the increase in the synchrotron seed photons becomes apparent.
Then the delay between the SSC emission and synchrotron emission may be less prominent,
as shown in \citet{chen_etal:2011:multizone_code_mrk421} for Mrk 421. 
We here investigate this scenario for PKS 0208-512 
with time-dependent simulations, in which $t_{acc}$ is reduced by a factor of 200 in a thin layer
without any change in B. The results are shown in 
Fig.\,\ref{fig:sscacc}.

In the time-dependent SSC model, the model parameteres are well constrained, 
and the resulting strength of magnetic field is only $0.4$mG in the quiescent state, 
making the cooling of the electrons quite slow compared to the
flare time scale. 
One result of this slow cooling is that any existing energized electrons, 
or injected new high-energy particle population 
\citep{barkov_2012:jetstar:749.119, zacharias_2012:nonlinearSSC},
will survive for a while without being able to cool down to the quiescent state soon.
This explains the electron distributions we see (Fig.\,\ref{fig:sscacc} bottom), 
where the electrons are accelerated to higher energy because of the perturbation, 
but take a long time to cool down afterwards. 
The resulting light curves (except radio and \xray) would show sharp rising time, 
but much longer decay time (Fig.\,\ref{fig:sscacc} middle). 
However, such phenomena are not seen in observations. 
Therefore, the SSC with electron distribution change scenario,
including the one in which the change is caused by the increase of particle acceleration efficiency, 
does not explain the flares in PKS 0208-512 neither.

\subsection{Dusty torus EC scenarios}
\label{ec}
Having established the difficulty to explain optical orphan flares and a lack of time delays in a pure SSC scenario, we now turn to external Compton (EC) 
processes as the main mechanism responsible for \gray emission. 
We chose the dusty torus as source of the soft photons.
The torus is assumed to emit black body radiation with a temperature of about 370 K
\citep{gg_tavecchio:2009:canonical_blazars}.
This external photon field is constant, if, as we do, one considers standing perturbations in the jet (for more discussion of this issue see \S\,\ref{dis:ext}).
The detection of FSRQs by imaging atmospheric-Cherenkov telescopes (IACT) above 100 GeV 
\citep{albert_etal:2008:3C279_magic_detection,alksic_2011:1222_magic_detection:730.8,hess_2013:1510} 
places the \gray emission site outside of the broad-line region (BLR), otherwise the \grays would suffer catastrophic
$\gamma-\gamma$ absorption with the BLR photons \citep{tavecchio_2012:flat_blr}.
Compared to the pure SSC scenario, the EC scenario has one additional parameter, 
namely the energy density of the 
external photons. We use $u_{ext}'$ to denote this energy density in the jet frame. This parameter is connected to the luminosity of the quasar thermal emission. However, the
large uncertainty in the torus radius and covering factor means it is poorly constrained. So in the EC
scenario, we fix the bulk Lorentz factor $\Gamma$ 
(hence $\delta$) to 40, which is the value used in the SSC scenarios and
is also close to the largest value determined in VLBI observation of quasar
jets \citep{jorstad_etal:2005:VLBA}.

\begin{figure}
\centering
%\vspace*{2cm}
\includegraphics[width=1.0\linewidth,height=0.87\linewidth,clip=]{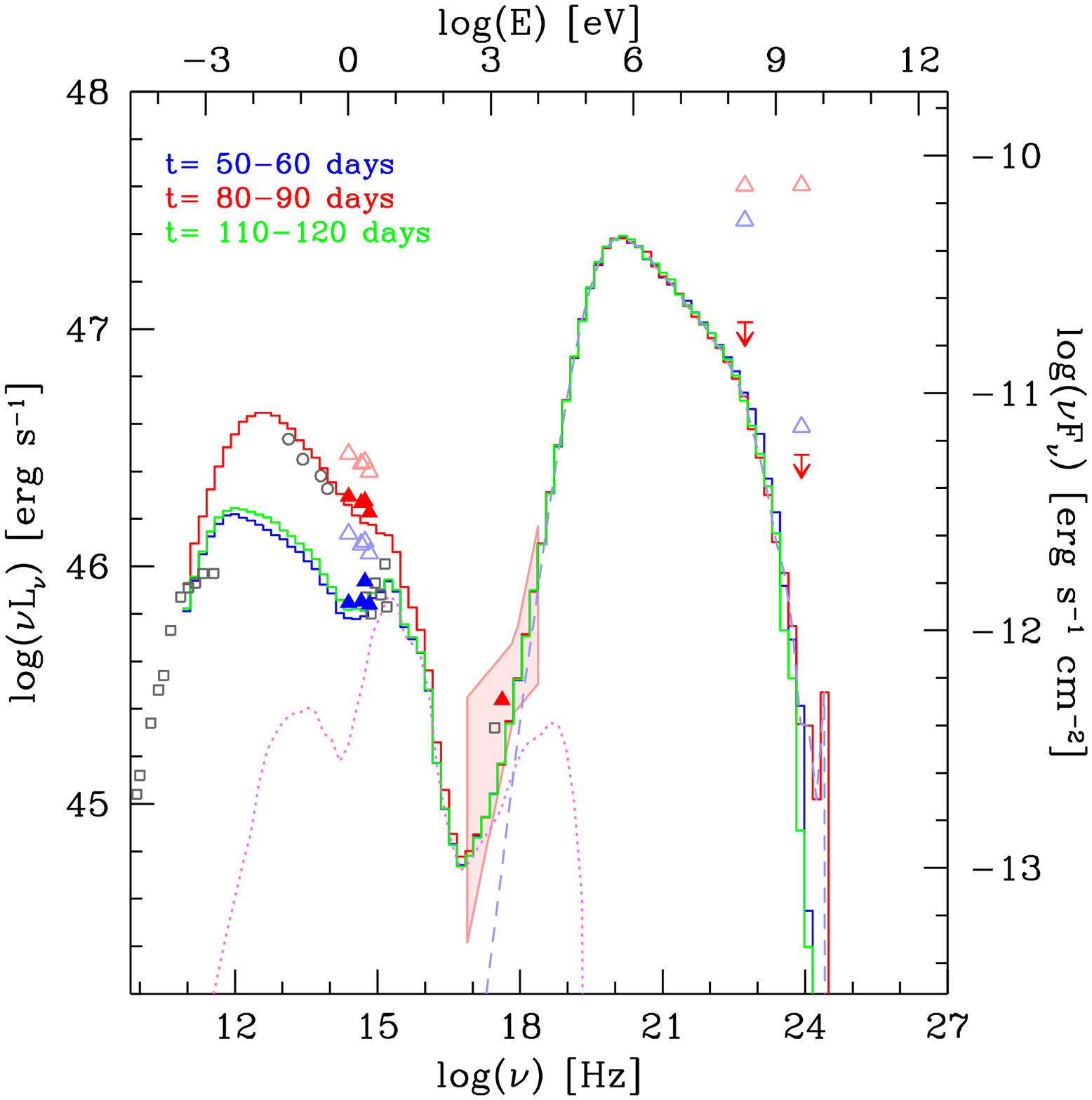}
\hfill
\includegraphics[width=1.0\linewidth,height=0.84\linewidth,clip=]{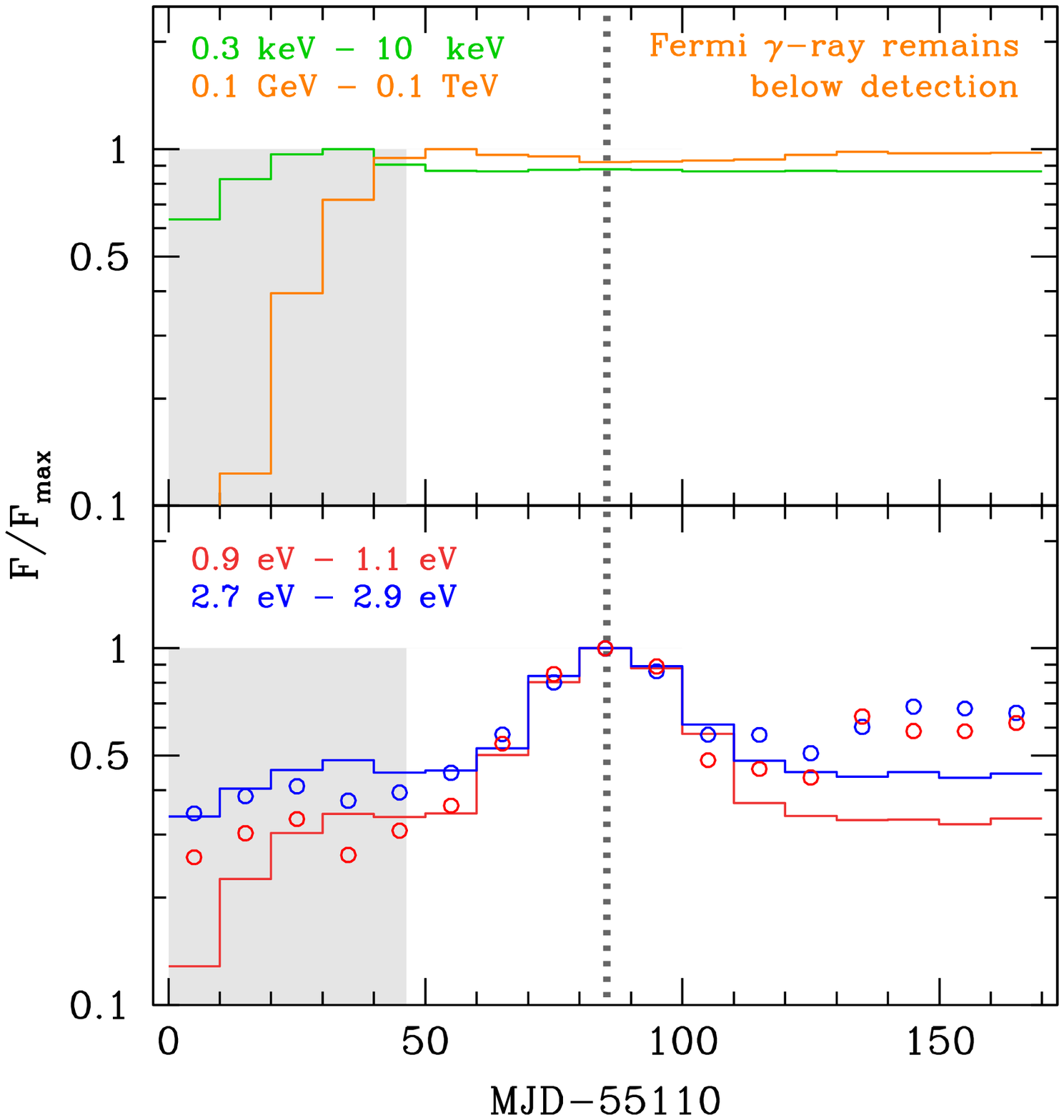}
\hfill
\includegraphics[width=1.0\linewidth,height=0.84\linewidth,clip=]{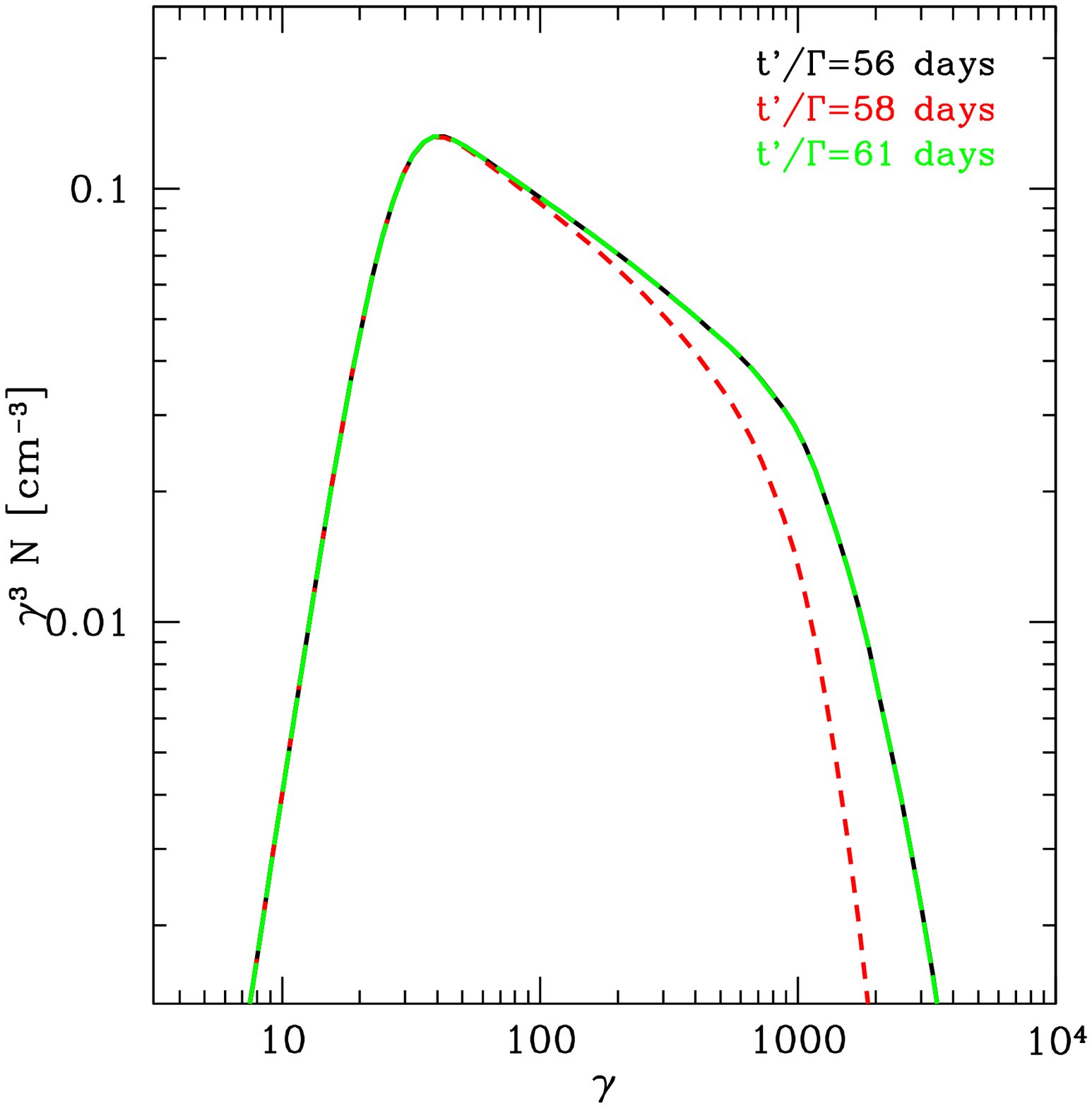}
\hfill
\caption{The SEDs (top), light curves (middle) and electron distributions (bottom) of the EC/dusty torus 
case with change of the magnetic field as the cause of the flare. 
The color schemes are similar to those in Fig.\,\ref{fig:sscb}, except for the EC component as blue dashed line in the SED. We also plot the SSC component, but its flux is below the flux scale shown here.}
\label{fig:ecb}       % Give a unique label
\end{figure}

\subsubsection{Brief change of magnetic-field strength}
\label{ec:B}

Fig.\,\ref{fig:ecb} shows results for the EC case with a brief change of magnetic-field strength as
the cause of the flare. The magnetic-field energy density is increased by a factor of 27 in the perturbation layer. 

Because stronger magnetic field imply stronger radiative cooling, 
the electrons change to a slightly lower energy state which, however, has minimal effects on the final radiation. 
The optical synchrotron emission becomes strongly variable, but the IC 
emissions remain quiet. 
This case is a fair reproduction of the observed orphan optical flares.
However, to spawn orphan optical flares and the optical/\gray
correlated flares in the same source,  
other processes must account for correlated flares.

\begin{table}
\centering
\caption{The parameters used for the quiescent state. The observation angle is always $1/\Gamma$ so that
the Doppler factor $\delta$ is equal to the bulk Lorentz factor $\Gamma$.
The volume radius $R=3Z/4$ in all cases. The time scale for acceleration and escape have the ratio $t_{acc}/t_{esc}=6.5$,
except during the flare caused by excess acceleration.}
\label{tab-1}       
\begin{tabular}{lll}
\hline
  & SSC & EC \\\hline
B(G) & $3.8\times10^{-4}$ & 2 \\
$\delta$ & 40 & 40 \\
$\gamma_{inj}$ & $2.4\times10^{3}$ & 20 \\
$Z(cm)$ & $2.4\times10^{18}$ & $2.4\times10^{18}$ \\
$n_e(cm^{-3})$ & 0.35 & $2\times10^{-4}$ \\
$t_{acc}(Z/c)$ & 120 & $9.75\times10^{-5}$ \\
$u_{ext}'(erg/cm^{3})$ & --- & 3.3 \\
\hline
\end{tabular}
% Or use
%\vspace*{5cm}  % with the correct table height
\end{table}

\subsubsection{Brief change of acceleration efficiency}
\label{ec:acc}

\begin{figure}
\centering
%\vspace*{2cm}
\includegraphics[width=1.0\linewidth,height=0.87\linewidth,clip=]{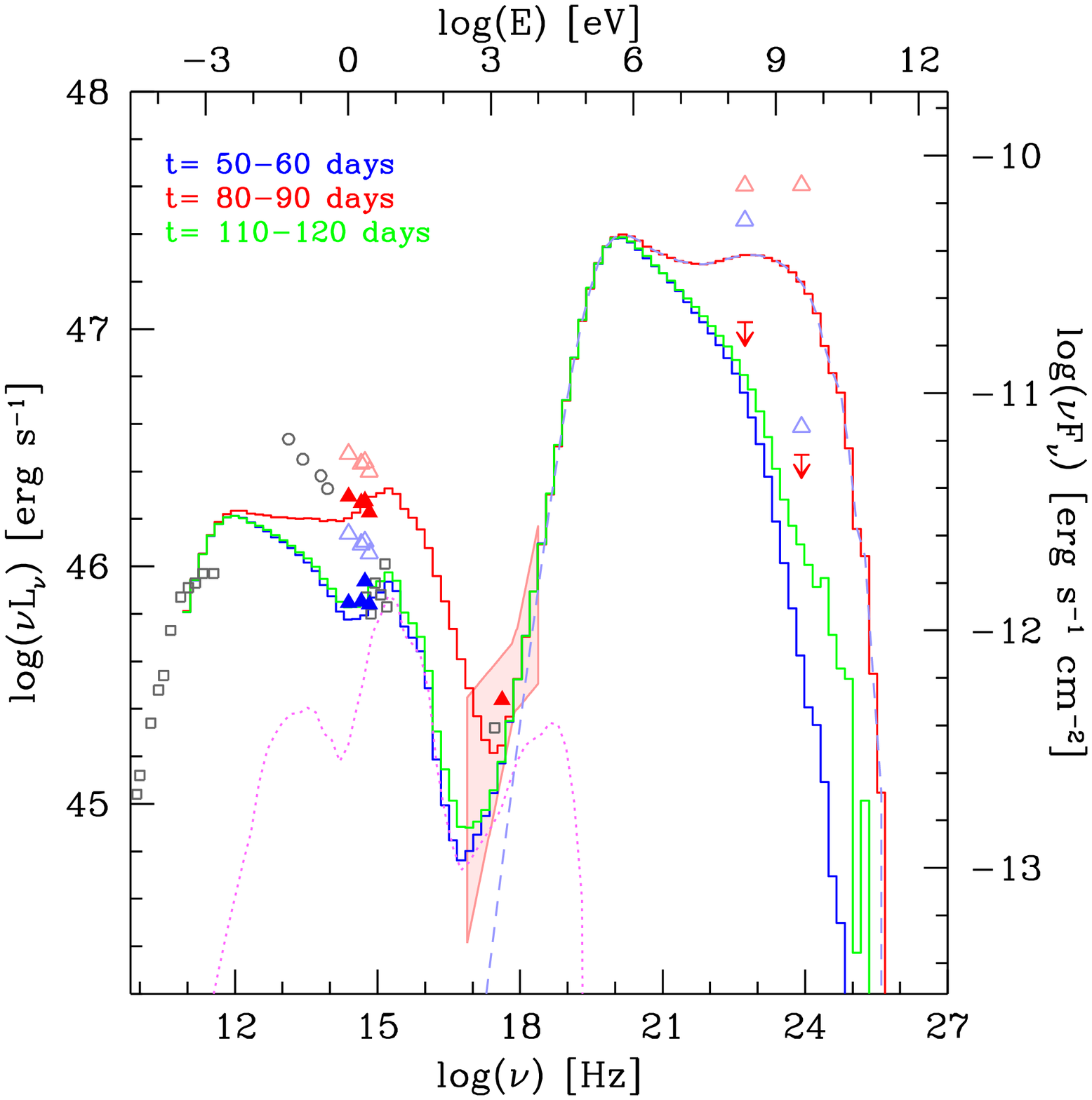}
\includegraphics[width=1.0\linewidth,height=0.84\linewidth,clip]{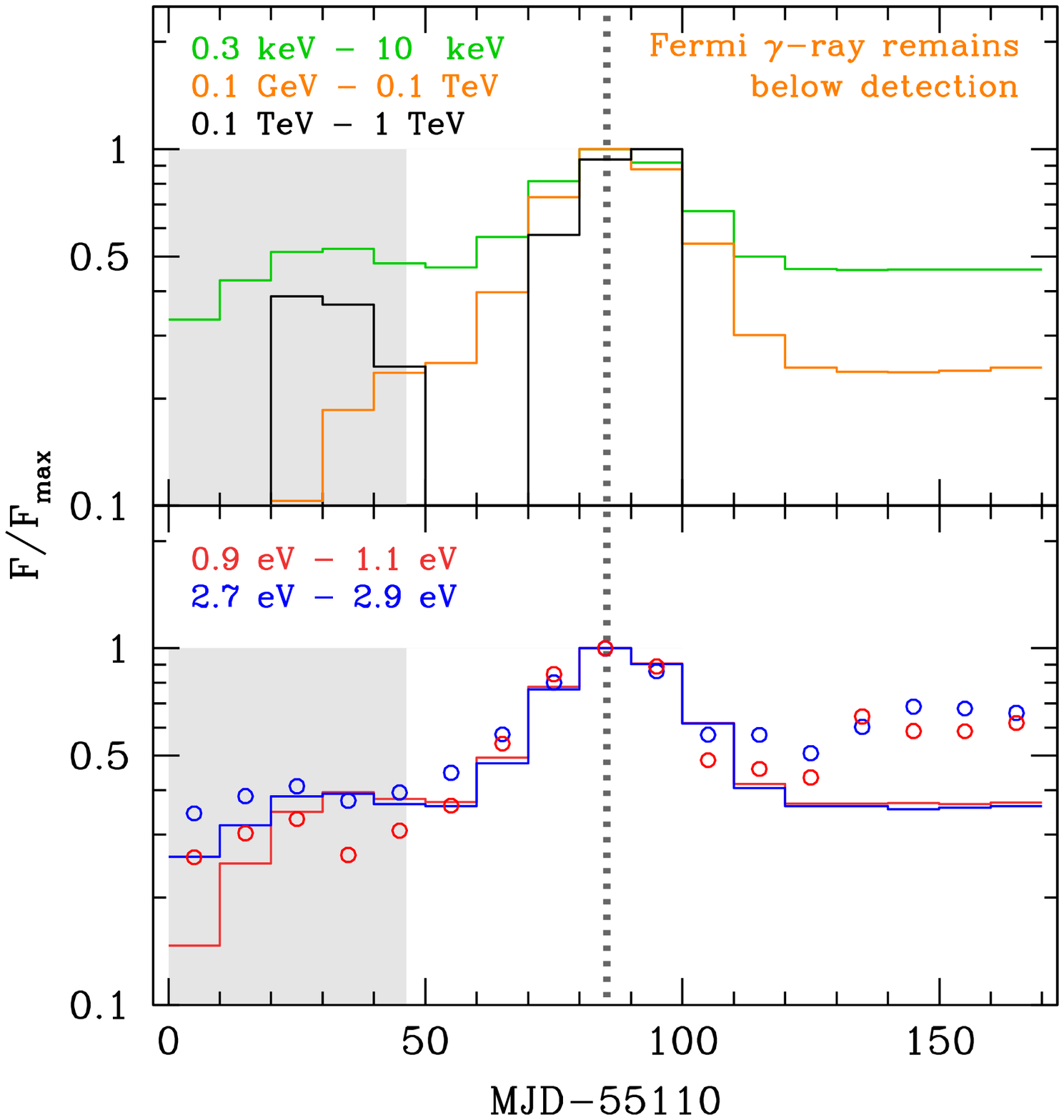}
\includegraphics[width=1.0\linewidth,height=0.84\linewidth,clip]{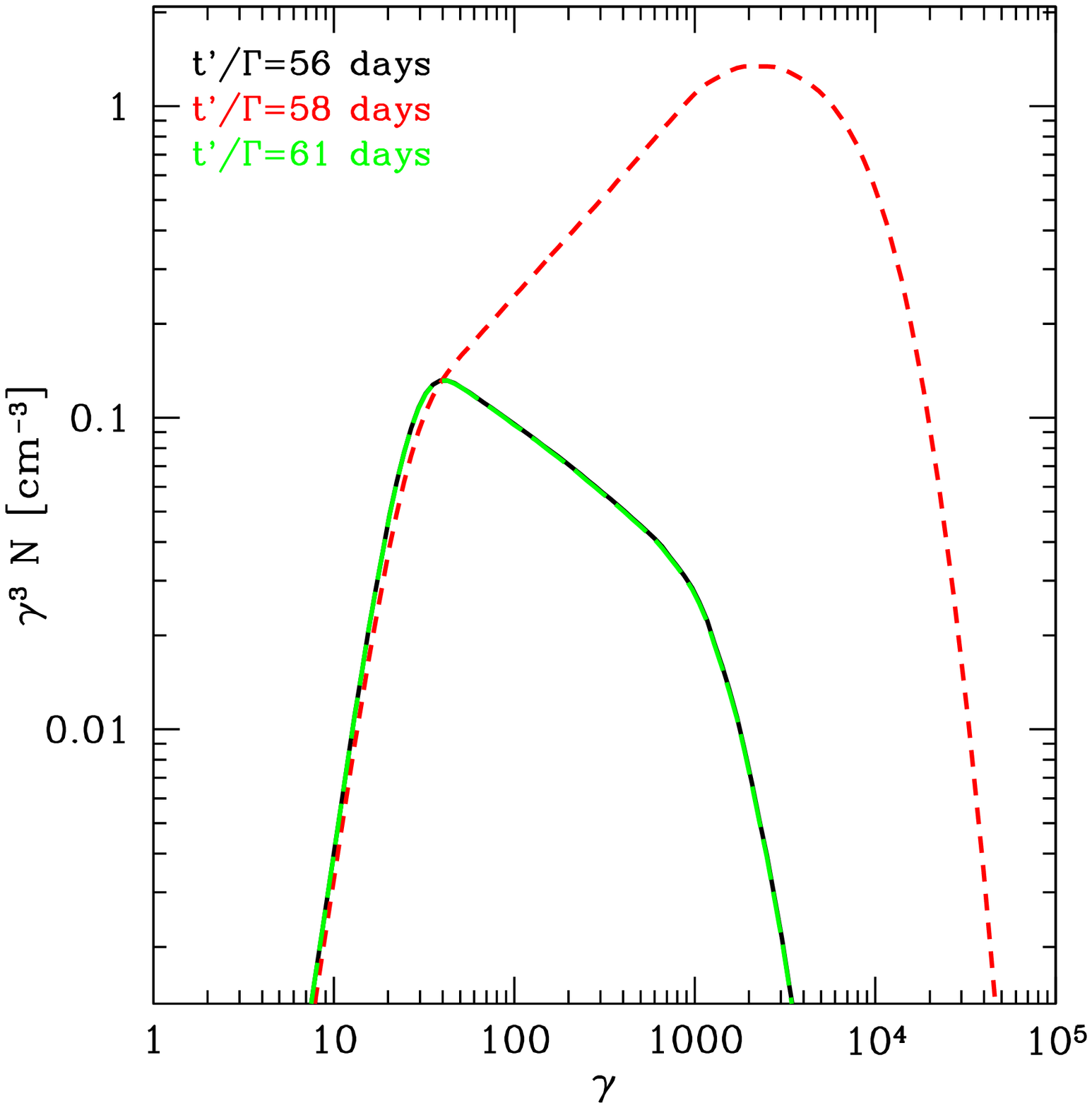}
\caption{The SEDs (top), light curves (middle) and electron distributions (bottom) of the EC/dusty torus 
case with change of particle-acceleration efficiency as cause of the flare. 
The color scheme is similar to that in Fig.\,\ref{fig:sscb}, 
except the blue dashed line in the SED shows the EC component.}
\label{fig:ecacc}       % Give a unique label
\end{figure}

Under the EC framework, we also evaluate the effect of a changing particle acceleration efficiency in a thin layer, in which $t_{acc}$ is reduced by a factor of 2.2 without change in $B$.
The results are shown in Fig.\,\ref{fig:ecacc}.

When the particle
acceleration efficiency increases, the electron distribution does not only become harder, but also reaches a higher cut-off
energy, albeit only for a short period of time (see Fig.\,\ref{fig:ecacc} bottom), 
which leads to multiple observable effects.

As seen in Fig.\,\ref{fig:ecacc}, the optical and \gray activities are well correlated, 
with no noticeable delay. This demonstrates that correlated optical/\gray flares 
are possible in the EC framework. 
In order to keep the consistency among different cases, we use the same quiescent state parameters that
are used for other cases, which matches the pre-flare SED with that of flare 2. We have also run simulations
with slightly different quiescent state parameters to match the SEDs of flare 1, which exhibits correlated optical/\gray flares.
The results do not show any
qualitive difference compared to the current case. Furthermore, the light curve of flare 1 is difficult to match accurately.
It appears to be a flare complex with multiple overlaping flaring elements, rather than a single isolated flare.

Another prominent feature of the results
is the spectral hardening during the flare, both in synchrotron and IC emission, which reflects the hardening of
the electron spectrum. This behaviour
is observed in \grays from \fermi monitored FSRQs
\citep{abdo_etal:2010:spectral_properties_of_blazars}. In the narrow optical and infrared band, 
observations usually reveal a redder-when-brighter trend \citep{bonning_2012:smarts_blazar}, 
most likely due to the contamination from the quasar thermal emission, whose influence is also seen in the simulated
light curves in Fig.\,\ref{fig:ecacc}. Observations in a wider far-infrared band can better
resolve the spectral behavior of the synchrotron component
\citep[see][who with \herschel observe a harder-when-brighter trend 
in PKS 1510-089]{nalewajko_2012:1510_herschel}.

The simulation indicates that the SED extends to higher energy during the flare, causing a 
significant VHE \gray flare on account of the higher maximum electron energy. For the same reason, the synchrotron
emission also extends to higher energy, causing an ultraviolet flare and a contribution of
synchrotron emission to 
the soft \xray band. The available simultaneous multi-wavelength data for PKS 0208-512 are insufficient 
to confirm the presence of these signals, but three other FSRQs have indeed been detected by 
Cherenkov Telescopes during flaring states 
\citep{albert_etal:2008:3C279_magic_detection,alksic_2011:1222_magic_detection:730.8,hess_2013:1510}. 

A soft-\xray excess was recently observed from one FSRQ (PMN J2345-1555) 
during a major flare,
accompanied by spectral hardening in GeV \grays \citep{gg_2013:2345_redblue}.
Our simulation suggests that \xray softening, ultraviolet/VHE \gray flaring, 
and far-infrared/\gray spectral hardening
are typical features of correlated optical/\gray flares. We suggest that ultraviolet flares and a strong spectral softening
in the \xray band can be used as triggers for IACT observations in search of VHE emission from FSRQs.
However, we note that even during strong flares of bright FSRQs, 
the predicted VHE flux is relatively low and persists only for
a short period of time. 
This flux is likely to be even lower if EBL absorption is taken into consideration. 
So the observation of FSRQs with existing IACTs may remain challenging.

\subsubsection{Brief change of magnetic field and acceleration}
We have also explored the scenario that magnetic field strength and particle acceleration are changed
simultaneously. The result reflects a mixture of the individual effects described in \S \ref{ec:B} and \S \ref{ec:acc}.
This mixture results in correlated optical/\gray flares, with the optical flare being stronger. 
The exact ratio of the flare amplitudes depends on the relative amplitude of the magnetic field/acceleration change.
Both the synchrotron and IC spectra harden and extend to higher energy, but not
as strongly as without magnetic-field amplification (\S \ref{ec:acc}), if the amplitude of the optical flare is still tuned to match the observation.

\subsection{Brief change of magnetic field orientation}
\label{ec:theta}

All results described above are based on the simplification that the magnetic fields are fully disordered.
These results should not change significantly, if the magnetic field is ordered and the field direction is constant in time.
However, there is evidence that the polarization angles of blazars change during flares 
\citep[\eg][]{larionov_2008:3C279.492.389, marscher_etal:2010:pks1510}, which suggests
changes in the intrinsic magnetic-field direction, which can potentially lead to blazar
flares as well. 

To study such scenarios, we first modified the MCFP code so that it can handle 
anisotropic synchrotron emission from ordered magnetic fields. We assume the magnetic fields to have helical
structure, which is supported by VLBI observation of radio jets \citep{osullivan_2009:helical_jet:393.429O}.
The angle dependent synchrotron emissivity is calculated in a way similar to those in \citet{jamil_2012:anisotropic}.
For a helical magnetic field which has no radial component, the poloidal magnetic flux must 
remain constant along the jet, to ensure flux conservation. 
In our configuration this means any change of the magnetic-field direction should be accompanied
by a change of magnetic-field strength in the azimuthal direction (see Fig.\,\ref{fig:geo} right).
An increase in the azimuthal field component may arise when the plasma 
passes through a compression, after which magnetic tension, reconnection, and other MHD processes relax the magnetic field to its original state. A specification of the details of this process is beyond the scope of the current paper, and we will only impose a localized enhancement in the toroidal field component and investigate its effect on the polarization of optical emission.

We calculate the time-dependent degree and angle of linear polarization
of the synchrotron emission with the 3-D multi-zone synchrotron-polarization code 3DPol developed by \citet{zhang_2014:3dpol}.
To account for LTTEs, each axisymmetric zone is cut into 120 slices 
in azimuth, for each of which the 3DPol code calculates the Stokes parameters with proper transformation and retardation to observer time. 

\begin{figure}
\centering
%\vspace*{2cm}
\includegraphics[width=1.0\linewidth,height=0.87\linewidth,clip=]{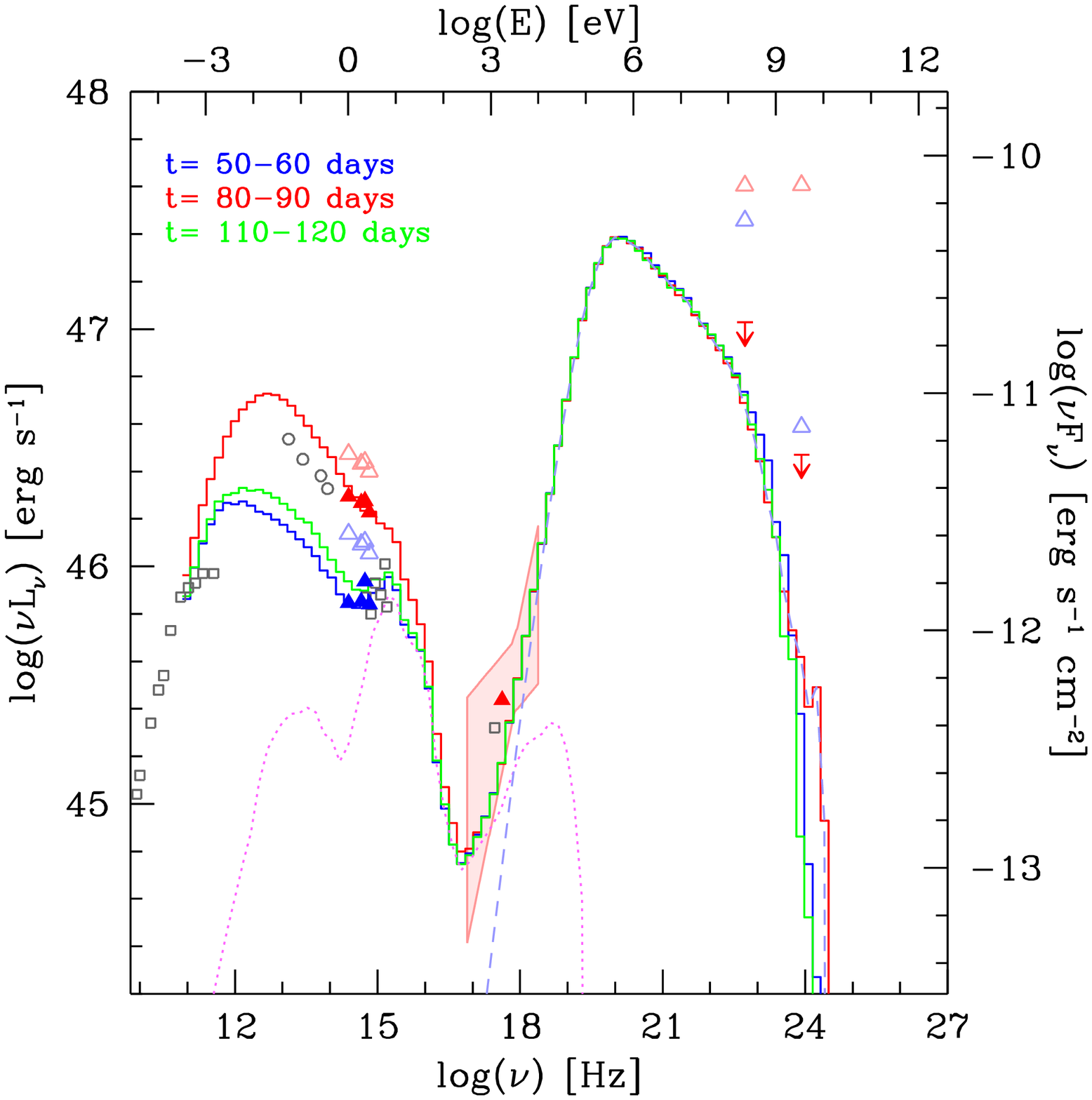}
\includegraphics[width=1.0\linewidth,height=0.87\linewidth,clip=]{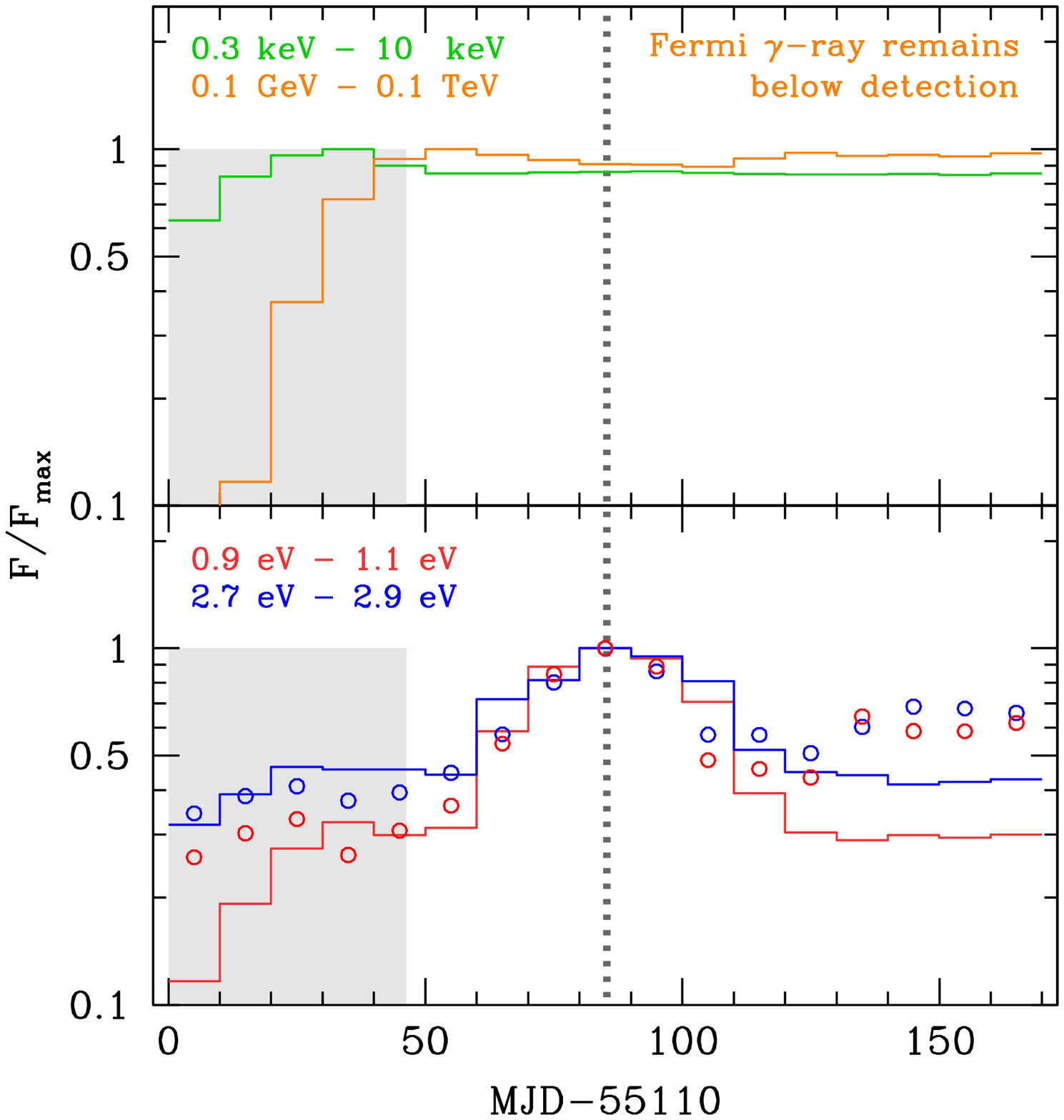}
\includegraphics[width=1.0\linewidth,height=0.87\linewidth,clip=]{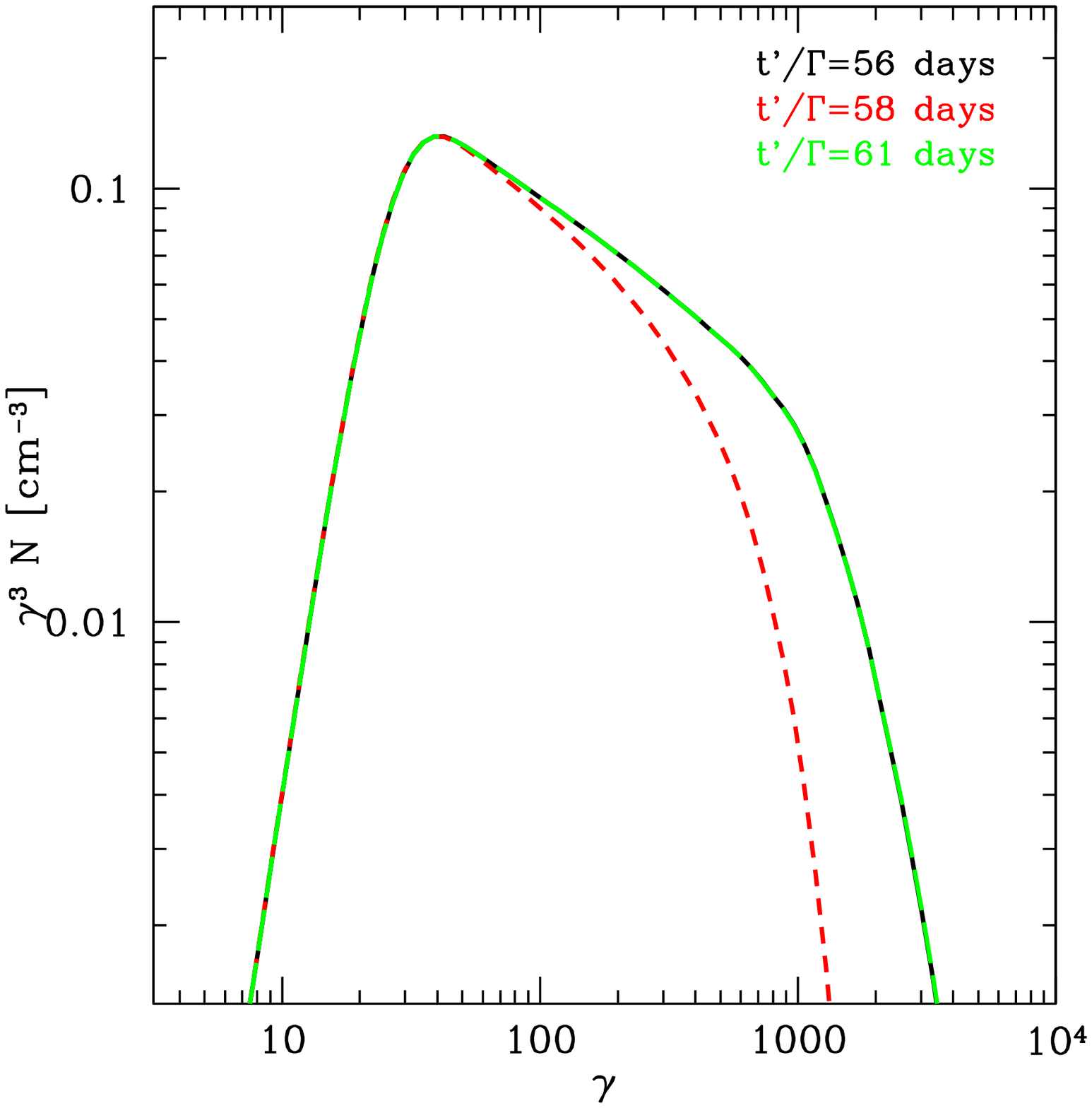}
\caption{The SEDs (top), light curves (middle) and electron distributions (bottom) of the EC/dusty torus
case with change of magnetic-field orientation as cause of the flare.
The color schemes are similar to those in Fig.\,\ref{fig:sscb}, 
except the blue dashed line in the SED shows the EC component.}
\label{fig:ect}       % Give a unique label
\end{figure}

We model flares linked to a change in magnetic-field orientation in the dusty torus EC framework.
The initial magnetic-field direction is at 45$^\circ$ to the jet axis (the poloidal and
azimuthal components of the field, $B_z$ and $B_\phi$, have equal strength), 
while the angle is 84.5$^\circ$ in a thin layer that relativistically propagates through the simulation volume.
This is realized by temporarily increasing $B_{\phi}$, while keeping $B_z$ and the acceleration unchanged.
Other parameters are identical to the other EC cases\footnote{This 
case is similar to the scenario 1 of \citet{zhang_2014:3dpol}, except for the choice of parameters.}.

\begin{figure*}
\centering
\includegraphics[width=0.485\linewidth]{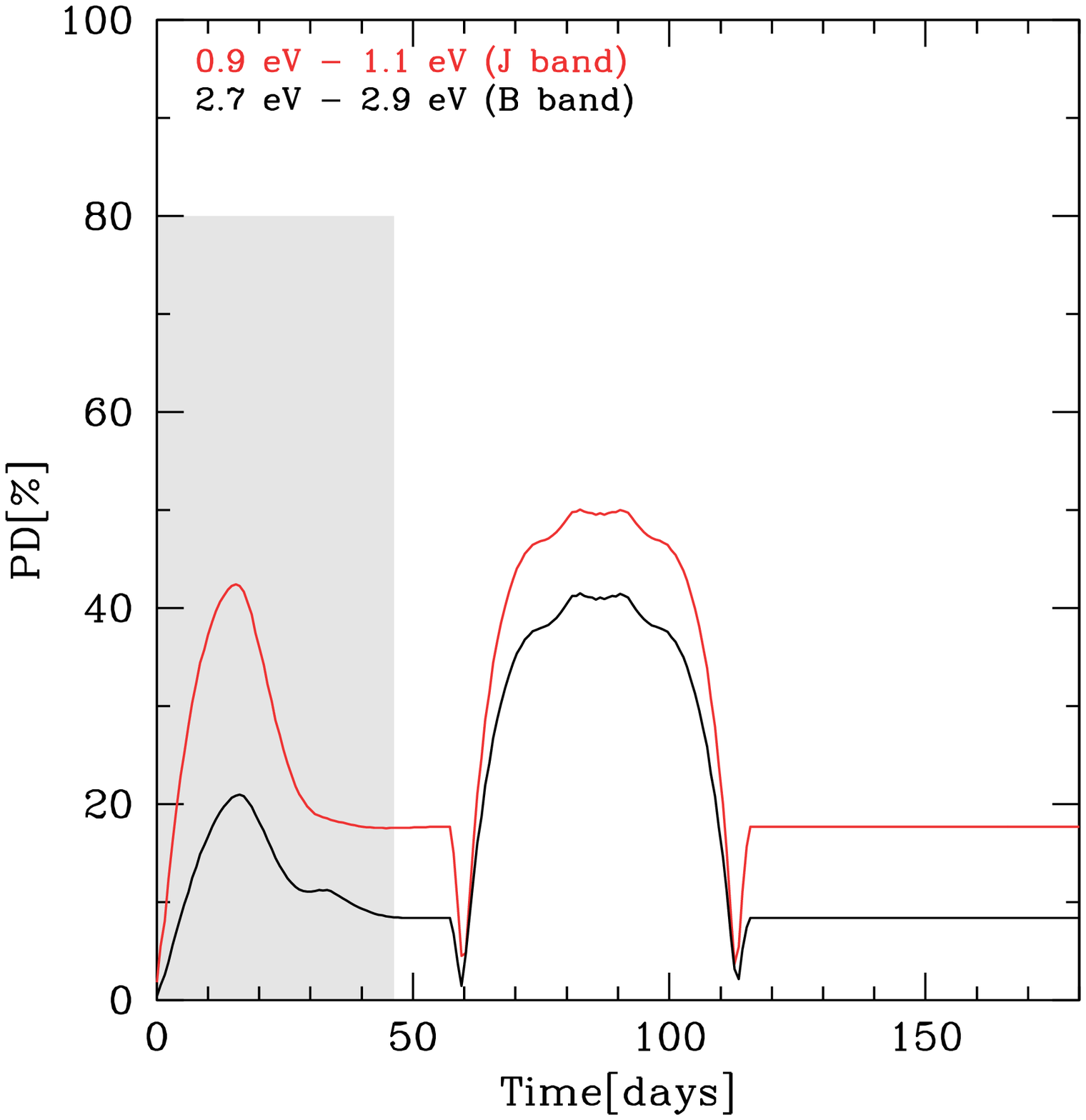}
\hfill
\includegraphics[width=0.485\linewidth]{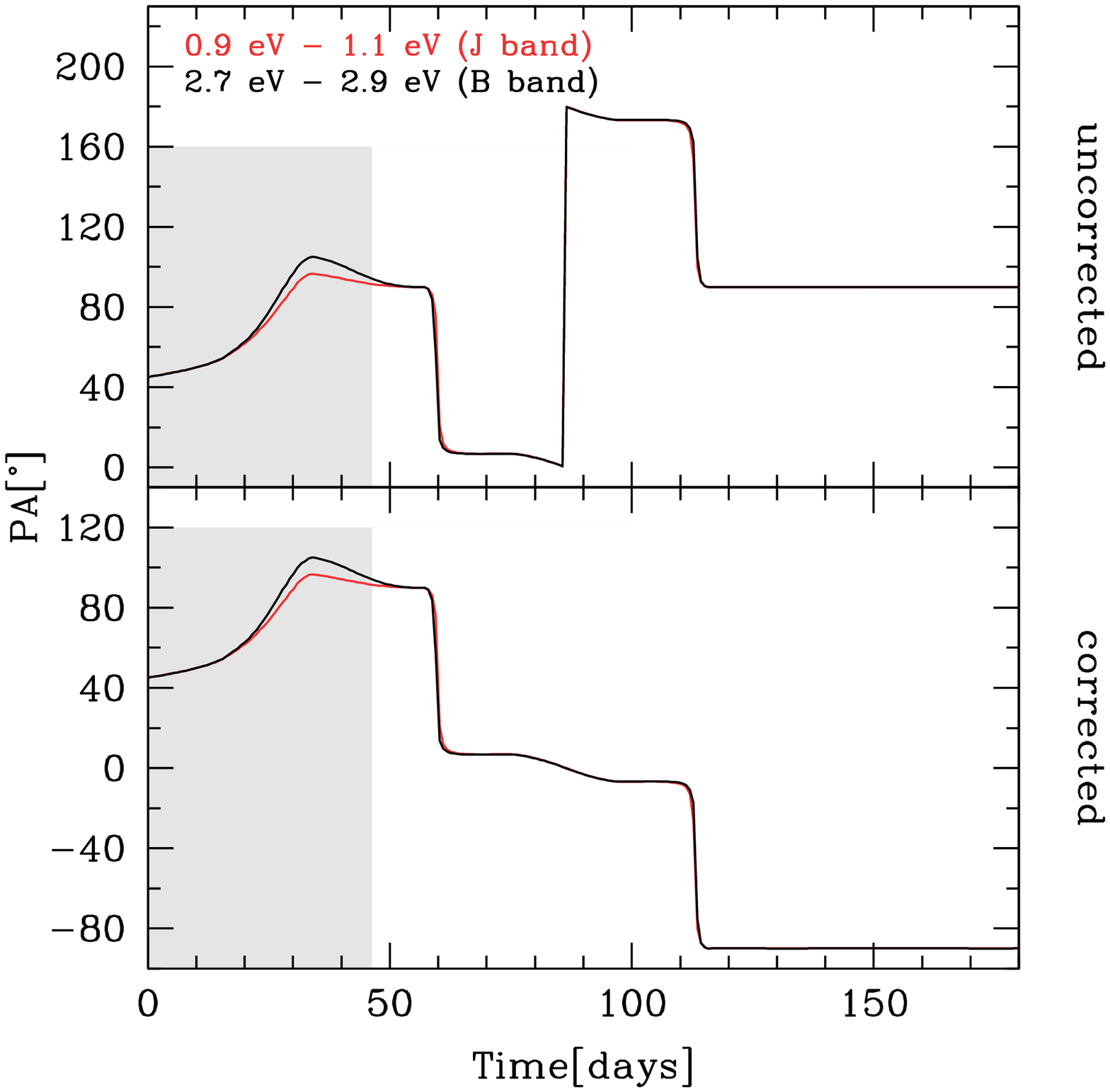}
\hfill
\caption{Synchrotron polarization degree (PD, left) and polarization angle (PA, right) 
for two optical bands in the EC/dusty torus scenario 
with change of magnetic-field orientation as cause of the flare. 
The shaded grey areas mark the phase when the simulation is still in setup phase. 
0$^\circ$ of the PA is set for its electric vector to be along the jet axis.
The uncorrected PAs (upper panel) are limited to
[0$^\circ$,180$^\circ$), while the corrected PAs (lower panel) assume that any two consective PA values 
change by less than 90$^\circ$.}
\label{fig:ect_pdpa}
\end{figure*}

The results are shown in
Fig.\,\ref{fig:ect} \& \ref{fig:ect_pdpa}.
Fig.\,\ref{fig:ect} appears similar to Fig.\,\ref{fig:ecb} at first glace, \ie, the case with change of the B-field strength, as both of them predict flares in the optical without \gray counterpart. The shape of the synchrotron flares is notably different, though. A variation in the magnetic-field direction yields longer-lasting high states, along with faster flux increase and decay.
%or, to turn the argument around, the intrinsic time scale of the jet perturbation is smaller for the same observable flare.

The difference in apparent flare shape arises from LTTEs: The emission from the lower near part of the cylinder will arrive first, 
and $B_\phi$ fully contributes to the observed synchrotron emission, because it is locally oriented 
perpendicular to the line of sight. As the perturbation moves along, we observe enhanced emission from the near part of the cylinder coincidentally with that from the central parts of the cylinder \citep[see illustration in Fig. 8 of][]{chen_etal:2011:multizone_code_mrk421}, where the increased $B_\phi$ does not contribute to the observed emission, because in a helical field configuration $B_\phi$ is aligned with the line of sight, when the observer is located 90$^\circ$ from the axis. Later, when we observe enhanced synchrotron emission from 
the upper rear part, the enhancement in $B_\phi$ is again reflected in the emissivity. 
This explains why the flare changes rapidly at the begining and end of the flare, while remains relatively stable around the peak.
%and so the flare appears longer lasting.

Fig.\,\ref{fig:ect_pdpa} shows polarization results consistent with those presented in \citet{zhang_2014:3dpol} for PKS 1510-089. The polarization degree shows a substantial increase during the flare, except at the beginning and end of the flare, 
when temporary decreases are seen. This can be understood by noticing that at the beginning and end 
$B_z$ dominates the polarization because it is always perpendicular to the line of sight, but around the peak of the flare 
$B_\phi$ dominates. The higher frequency band shows a smaller polarization degree
because of contamination with thermal emission (the big blue bump). 
We also notice small dips in the polarization degree near the peak of the flare that arise from beam and line-of-sight depolarization of emission from the central region of the simulation volume. 

Similar geometric effects cause the relevant PA changes shown in the right panel of Fig.\,\ref{fig:ect_pdpa}. During the
first half of the flare, the emission from the near side dominates, and so the polarization angle is close to
5.5$^\circ$ (for a right-handed helix). In the second half of the flare, the polarization angle
is close to 174.5$^\circ$. Before and after the flare, the emission from 
$B_z$ is stronger than that from $B_\phi$, and so the PA is 90$^\circ$. 

The lower panel of Fig.\,\ref{fig:ect_pdpa} right shows that, if we `correct' the PA by forcing it to always change by less than 90$^\circ$,
which is a common practice in the analysis of observations \citep[\eg][]{marscher_etal:2010:pks1510},
a single flare can be interpreted as
going through a PA rotation by 180$^\circ$. If the process repeats, causing multiple flares,
the total PA rotation can reach values much larger than 180$^\circ$. If the magnetic helix is left-handed,
the rotation will progress in an opposite direction.

\section{Discussion}
\label{discussion}
With our time dependent inhomogeneous blazar model, 
we studied in both the SSC and the EC framework FSRQ flares that are caused by changes in magnetic field and particle-acceleration efficiency.

SSC models with magnetic-field change predict correlated optical/\gray flares, 
with a slight delay of the \gray emission.
This delay is caused by the extra propagation time of synchrotron photon 
before the self-Compton scattering \citep{sokolov_marscher_mchardy:2004:SSC}. Such delays in FSRQs have not been confirmed by observations. SSC models with localized enhancement in the
acceleration efficiency do not reproduce the observations either, because they generally predict flux increases
without decay on the same time scale. 
In particular, the orphan optical flare observed in PKS 0208-512 does not find
a natural explanation in the SSC framework. All in all, the 
SSC scenario appears to be less favorable for FSRQs. 

Using the EC framework,
we reproduced both orphan optical and optical/\gray correlated flares with the same quiescent-state 
setup. In the simulation the difference lies in the cause of the flares, 
i.e. whether the magnetic field or particle-acceleration efficiency is changed. 
From an astrophysical point of view, this difference can be
tied to the allocation of free energy between magnetization and turbulence. 
Interestingly, this may depend on the 
initial orientation of the magnetic field.  

MHD simulations of magnetic-field amplification \citep{mizuno_2011:mhd_bampli,2014arXiv1401.7080M} 
show that when the magnetic field is perpendicular to the plasma flow, compression increases the share of energy that is transferred to the large-scale
magnetic field. 
If the magnetic field is parallel to the plasma flow, more energy is channelled into turbulent
magnetic field and possibly particle acceleration.
We speculate that the former case corresponds to the orphan optical flares, 
and the latter case corresponds to the optical/\gray correlated flares.

At the same time, the compression of perpendicular magnetic field should lead to a stronger optical polarization. In this interpretation,
an increase in polarization degree will coincide with an 
increase in optical flux while its correlation with \gray flux will be weak.
Whether such a correlation between the degree of polarization and the optical/\gray flux ratio
exists is yet to be observationally tested.

An enhancement in particle-acceleration efficiency
(possibly associated with parallel magnetic field and stronger turbulent magnetic-field amplification)
in the EC framework does not only explain the optical/\gray correlated flares,
but also produces interesting features that resemble those observed from many FSRQ flares.
These include a spectral hardening of infrared and \gray emission, 
a soft-\xray excess, and the rare detections of VHE \grays. All the above arise
because the enhanced particle acceleration will 
harden the electron spectrum and extend the
maximum Lorentz factor $\gamma_{max}$ at the same time. 
Synchrotron and IC emission will simply reflect the shape of the electron spectrum:
Its hardening manifests itself
as spectral hardening in both the synchrotron (infrared) and the IC (\gray) emission, 
while the larger maximum Lorentz factor leads to higher cut-off energies of 
the two emission components.
Our results also suggest
that activity in the ultraviolet band and the softness of the \xray spectrum can be important 
diagnostics of VHE flaring states, although VHE detection 
will probably remain difficult on account of the low flux even during flares and the short duration
of these flares.

\subsection{Ordered Magnetic Field and Polarization}
\label{dis:pol}

Using a simple helical magnetic field, we studied the effects of 
an ordered magnetic field on the multiwaveband variability and the optical polarization.
The most interesting finding is that this simple scenario produces not only an orphan-optical flare,
but also an increase in polarization degree and a perceived `rotation' of polarization angle (PA).
This `rotation' is in parts caused by LTTEs and hence does not constitute evidence of plasma moving in a helical trajectory along the
magnetic field lines of the jet \citep{marscher_etal:2010:pks1510}.
If there are multiple plasma perturbations, or the plasma crosses multiple external perturbations,
causing multiple flares, 
the measured PA rotation can be much larger than 180$^\circ$.
Interestingly,
S5 0716+714 has been observed to show a step-like rotation of PA \citep{ikejiri_2011:polarization:63.639},
which seems to be compatible with this scenario.

Our treatment of the compression of the magnetic helix and its later relaxation are highly simplified. Detailed studies and MHD simulations are needed to elucidate the details of the scenario. We also note that a combination of ordered and disordered
magnetic field must be expected and indeed appears required to
produce a correlation between polarization and optical/\gray flux ratio, as
we qualitively discussed earlier.

%Currently our model assumes the magnetic field is disordered, to simplify the calculation of 
%synchrotron emission.
%If the magnetic field has helical structure, the tightness of the helix affects 
%the observed synchrotron emission. If we are observing from approximately $\theta=1/\Gamma$ angle 
%(i.e. side-on $\theta'=90$ in the frame of the emitting blob), 
%this effect is relatively small, with the observed difference in flux between the most tightened helical 
%and purely longitudinal fields being $1/2$. 
%However, if we are observing the jet from face on direction ($\theta=0$),
%the purely longitudinal fields can result in zero observed synchrotron flux in the most extreme case.
%The effects of anisotropic synchrotron emission in a SSC scenario has been discussed by 
%\citep{jamil_2012:anisotropic} in a non-time dependent model. It appears possible that the change of 
%magnetic field orientation can cause orphan synchrotron flares. However, in the case of PKS 0208-512, the 
%problem of the SSC scenario is more in explaining the optical/\gray correlated flares rather than 
%the orphan optical flare.
%For the EC case, the change of field orientation has very similar effects as the change of field strength.
%So most of our results here should still apply.
%A more detailed analysis of the effect of ordered magnetic field in the time-dependent model is being
%undertaken.

\subsection{External Radiation Field}
\label{dis:ext}

Whereas this study only considers the dusty torus as source of external radiation, its results are qualitatively applicable to most other kinds of external photon sources. 
The size of the emission blob used in this study is comparable to the size of the dusty torus 
($R_{ir}=5\times10^{18}\ \mathrm{cm}$). If we did not consider a perturbation of the jet that is stationary in the host-galaxy frame, this perturbation would
travel a significant distance during a single flare, and consequently would
experience a considerable variation of the external photon density.
We can expect that perturbations traveling down the jet can cause a decrease in the \gray emission,
%variations which do not have optical counterparts, 
on account of the changes in the external radiation field \citep{dermer_etal92,dermer_schlickeiser:2002:transformation_external_fields},
while the optical emission remains steady.

\subsection{Particle Escape}
\label{dis:esc}

The rate of particle escape required for PKS 0208-512 in the EC scenario is very high.
The very efficient radiative cooling mandates that the acceleration time be $t_{acc}\sim 10^{-4}R/c$, otherwise particles cannot be accelerated to $\gamma\sim 1000$.
The spectral index of the particle distribution on the other hand requires 
the ratio $t_{acc}/t_{esc}$ to be about 6.5. Therefore $t_{esc}$ needs to be as short as $1.5\times 10^{-5}R/c$.
This particle escape is too fast to be explained by 
particle streaming out of the emission region. 

One possible explanation is that the particles are
accelerated in much smaller turbulent cells 
\citep[see][for discussion of the idea of turbulent cells]{marscher_2013:turb_model}. So the `escape' used in our model does not represent
escape from the emission region, but rather escape from the accelerators. 
For example, if the size of the turbulent cells is $R_{cell}=10^{-5}R/c$,
then $t_{esc}=1.5R_{cell}/c$. An ensemble of such small acceleration zones might exist in the jet, whose large-scale distribution may be planar as assumed in this work or follow a different geometry.
In this interpretation, the escaped particles are no longer being accelerated, 
but still contained in the jet, and would thus still contribute to both synchrotron and IC emission. Our treatment of particle and radiation spectra does not properly account for this possibility.
Instead, we ignore the escaped particles and effectively model the emission behavior of the particles in the accelerators.

In fact, particle escape from jet emission zones is a widespread but questionable concept, because at least IC scattering of ambient radiation is unavoidable and
the emission from the escaped particles may well dominate over those residing in the accelerator region. In this work we do not directly model the escaped particle population, because  simplifications such as the energy independence of the acceleration time scale or the constant Lorentz factor of the jets (i.e., the absence of internal shocks in the model) may be as influental in driving the parameters. 

Here, our emphasis was on exploring the impact of LTTEs. A forthcoming paper shall investigate small-scale acceleration zones embedded in a larger jet. For that purpose we shall add diffusive transport to the code to self-consistently model the three-dimensional expansion of a cloud of flare electrons that are produced in a small volume. As the geometry of the emission region is a critical parameter in modeling blazar flares \citep{sokolov_marscher_mchardy:2004:SSC,sokolov_marscher:2005:EC}, we may find that a different particle escape rate is needed to maintain the same characteristics of the SED. As this time, all parameters should be taken \emph{cum grano salis}.

\section{Conclusions}
\label{conclusion}

Using our 2-D Monte-Carlo/Fokker-Planck code, we have modeled FSRQ flares with several scenarios related to 
magnetic-field amplifications. Changes in the magnetic-field strength, its orientation,
and the particle-acceleration efficiency have been explored as causes of flares, with either the SSC or the EC mechanism
producing the \gray emission.

We conclude this paper with the following findings regarding FSRQs:

\begin{enumerate}
\item 
SSC scenarios are disfavored because they do not naturally produce orphan-optical flares,
and even for optical/\gray correlated flares, the model either predicts a delay of \gray flare, 
or the flux would not subside after the flare.
\item
The difference between orphan optical and optical/\gray correlated flares can be understood as being caused by
different initial orientation of the magnetic field in the plasma and subsequent effects
on magnetic-field amplification.
\item
Changes in the particle-acceleration efficiency in an EC model can explain several 
features observed during flares, including:
\begin{itemize}
\item spectral hardening of infrared and \grays;
\item \xray soft-excess;
\item rare detections of VHE \grays.
\end{itemize}
\item
A correlation may exist between optical polarization degree and optical/\gray flux ratio.
\item
The polarization degree change and polarization angle rotation of $\geq 180^\circ$ can be explained by 
compression and subsequent relaxation of helical magnetic fields in combination with light-travel-time effects.
\end{enumerate}

\section*{Acknowledgements}

The authors thank A. Barnacka for useful discussions. 
XC and MP acknowledge support by the Helmholtz Alliance for Astroparticle Physics HAP funded by the Initiative and Networking Fund of the Helmholtz Association. 
HZ acknowledges supports by NASA through Fermi Guest Investigator Grant no. NNX12AP20G, and by the LANL/LDRD program and by DoE/Office of Fusion Energy Science through CMSO. 
GF acknowledges support by NASA grant NNX12AE43G. 
MB acknowledges support through the South African Research Chair Initiative of the National Research Foundation and the Department of Science and Technology of South Africa.

% BibTeX or Biber users please use (the style is already called in the class, ensure that the "woc.bst" style is in your local directory)

\bibliography{refs_all}
%\begin{thebibliography}{}
% Journal Author, Journal \textbf{Volume}, page numbers (year)
%Book Author, \textit{Book title} (Publisher, place, year) page numbers
% etc
% \end{thebibliography}
\end{document}